\documentclass[preprint,12pt, 
aps,amsmath,amssymb,nofootinbib,superscriptaddress,hyperref,
]{revtex4-1} 
\usepackage[utf8]{inputenc}
\usepackage{epsfig} 
\usepackage{wasysym} 
\usepackage{mathrsfs} 
\usepackage{graphicx} 
\usepackage{amsfonts} 
\usepackage{amsbsy} 
\usepackage{amscd} 
\usepackage{pstricks} 
\usepackage{multirow}
\usepackage{slashed} 
\usepackage{tikz}
\usetikzlibrary{decorations.pathmorphing,decorations.markings}
\newcommand{\beq}{\begin{equation}}
\newcommand{\eeq}{\end{equation}}
\newcommand{\bea}{\begin{eqnarray}}
\newcommand{\eea}{\end{eqnarray}}
\newcommand{\beas}{\begin{eqnarray*}}
\newcommand{\eeas}{\end{eqnarray*}}
\newcommand{\bi}{\begin{itemize}}
\newcommand{\ei}{\end{itemize}}

\def\tev{\,{\ifmmode\mathrm {TeV}\else TeV\fi}}
\def\gev{\,{\ifmmode\mathrm {GeV}\else GeV\fi}}
\def\to{\rightarrow}

\allowdisplaybreaks

\begin{document}
\preprint{FERMILAB-PUB-19-268-T}

\title{
Long-lived TeV-scale right-handed neutrino production 
at the LHC in gauged $U(1)_X$ model
}

\author{Arindam Das}
\email{arindam.das@het.phys.sci.osaka-u.ac.jp}
\affiliation{Department of Physics, Osaka University, Toyonaka, Osaka 560-0043, Japan}

\author{P. S. Bhupal Dev}
\email{bdev@wustl.edu}
\affiliation{Department of Physics and McDonnell Center for the Space Sciences, Washington University, St. Louis, MO 63130, USA}
\affiliation{Theoretical Physics Department, Fermi National Accelerator Laboratory, P.O. Box 500, Batavia,
IL 60510, USA}

\author{Nobuchika Okada}
\email{okadan@ua.edu}
\affiliation{Department of Physics and Astronomy, University of Alabama, 
Tuscaloosa,  AL 35487, USA} 

\begin{abstract}

A gauged $U(1)_X$ extension of the Standard Model is a simple and consistent framework to naturally incorporate 
   three right-handed neutrinos (RHNs) for generating the observed light neutrino masses and mixing by the type-I seesaw mechanism. We examine the collider testability of the $U(1)_X$ model, both in its minimal form with the conventional charges, as well as with an alternative charge assignment, 
   via the resonant production of the $U(1)_X$ gauge boson ($Z^\prime$)  and its subsequent decay into a pair of RHNs. We first derive an updated upper limit on the new gauge coupling $g_X$ as a function of the $Z'$-boson mass from the latest LHC dilepton searches. Then we identify the maximum possible cross section 
   for the RHN pair-production under these constraints. Finally, we investigate the possibility of having one of the RHNs long-lived, even for a TeV-scale mass. Employing the general parametrization for the light neutrino mass matrix to reproduce the observed neutrino oscillation data, 
  we perform a parameter scan and find a simple formula for the maximum RHN lifetime  
  as a function of the lightest neutrino mass eigenvalue ($m_{\rm lightest}$). We find that for $m_{\rm lightest}\lesssim 10^{-5}$ eV, one of the RHNs in the minimal $U(1)_X$ scenario can be long-lived with a displaced-vertex signature 
  which can be searched for at the LHC and/or with a dedicated long-lived particle detector, such as MATHUSLA. In other words, once a long-lived RHN is observed, we can set an upper bound on the lightest neutrino mass in this model. 

\end{abstract}

\maketitle

\section{Introduction} 

The existence of nonzero neutrino masses and flavor mixing has been unambiguously established 
   by the observation of neutrino oscillation phenomena in solar, atmospheric, reactor and accelerator neutrinos~\cite{Tanabashi:2018oca}. 
This necessitates some extension of the Standard Model (SM) to incorporate 
   the neutrino mass matrix. 
Arguably, the simplest paradigm for explaining neutrino masses at tree-level is the type-I seesaw mechanism~\cite{Minkowski:1977sc, Mohapatra:1979ia, Yanagida:1979as, GellMann:1980vs, Schechter:1980gr}, 
   where SM-singlet right-handed neutrinos (RHNs) with  Majorana masses are introduced.

In the minimal seesaw extension of the SM, the RHNs are just put in `by hand'. In this phenomenological bottom-up approach, the mass scale of the RHNs is undetermined and can lie anywhere from the eV-scale to the Grand Unified Theory (GUT) scale~\cite{Drewes:2013gca}. From the experimental point of view, it is of particular interest if the RHN mass-scale is around TeV or smaller, so that they are kinematically accessible at the Large Hadron Collider (LHC) and can lead to the smoking-gun lepton number violating (LNV) signature of same-sign dilepton plus jets without missing transverse energy~\cite{Keung:1983uu}.  However, being SM-singlets, the RHNs in the minimal seesaw can be produced only through their mixing with the active neutrinos, which is generically required to be small for TeV-scale RHN masses in order to reproduce the observed light neutrino oscillation data~\cite{Ibarra:2010xw, Fernandez-Martinez:2016lgt, Das:2017nvm}, thereby suppressing the LNV signal from the RHNs at the LHC~\cite{Datta:1993nm, Panella:2001wq, Han:2006ip, delAguila:2007qnc, Dev:2013wba, Alva:2014gxa, Das:2015toa, Das:2016hof, Pascoli:2018heg}. For reviews on collider tests of RHNs, see e.g. Refs.~\cite{delAguila:2008cj, Atre:2009rg, Deppisch:2015qwa, Antusch:2016ejd, Cai:2017mow, Das:2018hph}, and for recent LHC searches, see Refs.~\cite{Aad:2015xaa, Sirunyan:2018xiv}.

A more compelling scenario, both theoretically and experimentally, is the  so-called minimal $B-L$ (baryon minus lepton number) model~\cite{Mohapatra:1980qe, Marshak:1979fm, Wetterich:1981bx, Masiero:1982fi, Buchmuller:1991ce}, 
   which is a gauge extension of the global $U(1)_{B-L}$ symmetry in the SM. 
In this model, the presence of three RHNs is essential to cancel the gauge and the mixed gauge-gravitational anomalies. 
Associated with the $U(1)_{B-L}$ gauge symmetry breaking, the RHNs acquire their Majorana masses, 
   and the SM neutrino masses are generated through the seesaw mechanism 
   after the electroweak symmetry breaking.  This model provides a new mechanism for the pair-production of RHNs at the LHC through the resonant production of the $B-L$ gauge boson $Z'$, which couples to the SM fermions as well as the RHNs. Therefore, the RHN production rate in this case is no longer suppressed by the RHN mixing with the active neutrinos, but only by the $Z'$ mass. In the context of the $B-L$ models, the prospect of discovering RHNs 
  at the LHC has been explored in Refs.~\cite{Basso:2008iv, Deppisch:2013cya, Kang:2015uoc, Cox:2017eme, Accomando:2017qcs}. This is however limited, even at the high-luminosity phase of the LHC (HL-LHC) due to the severe constraints on the $Z'$ mass from LHC dilepton searches~\cite{Sirunyan:2018exx, CMS:2018wsn, ATLAS:2019}, which restrict $m_{Z'}\gtrsim 4.5$ TeV.

The stringent dilepton bounds on the $Z'$ mass can in principle be relaxed, if we generalize the $B-L$ model to the so-called non-exotic $U(1)_X$ model~\cite{Appelquist:2002mw}, 
   where the particle content of the model is the same as the $B-L$ model,  
   while the $U(1)_X$ charge of each SM particle is defined as a linear combination
   of its $U(1)_Y$ hypercharge and $U(1)_{B-L}$ charge. 
In the presence of three RHNs, this model is also anomaly-free and 
   the seesaw mechanism is implemented after the breaking of the $U(1)_X$ and the electroweak symmetries. It has been shown~\cite{Das:2017flq, Das:2017deo} that with a suitable choice of the $U(1)_X$ charges, this model can realize a significant enhancement of the $Z'$ to RHN-pair branching ratio with respect to the dilepton branching ratio, thereby increasing the RHN discovery prospects at the LHC. 

In this paper, we examine the displaced vertex prospects of the RHNs in the $Z'$-induced resonant RHN pair-production channel at the LHC in the $U(1)_X$ model. For related discussion on displaced-vertex signatures of RHNs, see e.g. Refs.~\cite{Helo:2013esa, Izaguirre:2015pga, Alekhin:2015byh, Gago:2015vma, Antusch:2016vyf, Caputo:2017pit, Antusch:2017hhu, Dube:2017jgo, Cottin:2018kmq, Kling:2018wct, Cottin:2018nms, Abada:2018sfh, SHiP:2018xqw, Dercks:2018wum, Dib:2019ztn, Drewes:2019fou, Bondarenko:2019tss, Liu:2019ayx} in the context of minimal seesaw and \cite{Accomando:2016rpc, Deppisch:2018eth, Jana:2018rdf, Das:2018tbd, Deppisch:2019kvs} in the context of $U(1)_{B-L}$. For recent LHC searches relevant for long-lived RHNs decaying into final states containing two leptons, see Refs.~\cite{Sirunyan:2018mtv, Aad:2019kiz}. We consider two scenarios for the $B-L$ charges of the RHNs in the $U(1)_X$ model, both of which are consistent with anomaly cancellation: (i) the conventional case with a $B - L$ charge $-1$ assigned to all three RHNs, and (ii) the alternative case with a $B - L$ charge $-4$ assigned to two of the RHNs and +5 for the third one. In both cases, we present the dilepton production cross section from the $Z^\prime$ boson resonance  as a function of the U(1)$_X$ gauge coupling ($g_X$) and $Z^\prime$ boson mass ($m_{Z^\prime}$),  
  from which we can easily read off the upper (lower) bound   on $g_X$ ($m_{Z^\prime}$) for a fixed $m_{Z^\prime}$ ($g_X$) value  
  by using the current/future LHC constraints.  
We also present the RHN pair-production cross section from the $Z^\prime$ boson resonance 
  as a function of $g_X$ and $m_{Z^\prime}$, 
  so that we can find its maximum cross section once the LHC constraints on $g_X$ and $m_{Z^\prime}$ 
  are identified.  Once produced, each RHN will primarily decay into a pair of same-sign leptons plus jets through its suppressed coupling to the SM $W$ boson induced via its mixing with the active neutrinos.  
  
The main purpose of this paper is to show that in the $U(1)_X$ model, one of the three RHNs produced at the LHC can be long-lived even if its mass lies at the TeV scale. Employing the general parametrization of the neutrino mass matrix to reproduce the neutrino oscillation data, we perform a parameter scan and find the maximum RHN decay length as a function of the lightest neutrino mass eigenvalue for a given RHN mass. For a suitable range of the lightest neutrino mass, we find that one of the RHNs can be long-lived, to be explored by a displaced vertex search 
  either at the High-Luminosity LHC~\cite{Alimena:2019zri, talk} or dedicated long-lived particle search experiments using the LHC energy, such as CODEX-b~\cite{Gligorov:2017nwh}, MATHUSLA~\cite{Curtin:2018mvb}, AL3X~\cite{Gligorov:2018vkc}, FASER~\cite{Ariga:2018uku} and MAPP/MALL~\cite{Pinfold:2019nqj}. Thus, if such a long-lived RHN is ever observed in one of these experiments, that would indirectly probe the lightest active neutrino mass, which is one of the fundamental unknown parameters in the neutrino sector.

The rest of this paper is organized as follows: 
In Section~\ref{Sec2}, we briefly review the $U(1)_X$ model in two scenarios: (i)  the conventional case with a universal $U(1)_X$ charge for all three RHNs, and (ii) the alternative case with non-universal $U(1)_X$ charges assigned to three RHNs.  
In Section~\ref{Sec3}, we calculate the dilepton and RHN pair-production cross sections at the LHC  
   from the $Z^\prime$ boson resonance 
   as a function of the U(1)$_X$ gauge coupling and the $Z^\prime$ boson mass.  
The LHC constraints on these parameters will be read off from the results. 
In Section~\ref{Sec4}, we analyze the RHN decay length by using the general parametrization of the neutrino mass matrix. 
Performing a general parameter scan, we find a simple formula for the maximum RHN decay length and identify its correlation with the lightest neutrino mass. Our conclusions are given in 
Section~\ref{Sec5}.

\section{Gauged $U(1)$ extension of the Standard Model}
\label{Sec2}

We consider a simple gauged $U(1)$ extension of the SM based on the gauge group  
  $SU(3)_c \times SU(2)_L \times U(1)_Y \times U(1)_X$~\cite{Mohapatra:1980qe, Marshak:1979fm, Wetterich:1981bx, Masiero:1982fi, Buchmuller:1991ce} , where three RHNs are introduced to cancel 
  all the gauge and the mixed gauge-gravitational anomalies. We study two scenarios within this model: {\bf Case-I}  
is the conventional (minimal) U(1)$_X$ model~\cite{Appelquist:2002mw}, 
   which is a generalization of the minimal $B-L$ model 
   with the U(1)$_X$ charge assignment for a SM particle defined as a linear combination of 
   its $U(1)_Y$ hypercharge and $B-L$ charge. 
Except for the U(1)$_X$ charge assignments, the particle content is exactly the same 
   as the minimal $B-L$ model and all the RHNs are assigned a universal $U(1)_X$ charge. {\bf Case-II} 
 is what we call the ``alternative U(1)$_X$ model'', which is the only other type of $U(1)_X$ model we are aware of where all the gauge and the mixed gauge-gravitational anomalies are canceled. Although the fermion particle content of this model is the same as the minimal U(1)$_X$ model, 
  the charge assignment for RHNs is non-universal. Below we discuss each case in some details: 
\subsection{Case-I: Minimal $U(1)_X$ Model}
\begin{table}[t]
\begin{center}
\begin{tabular}{c|c|c|c|c}
\hline\hline
      &  $SU(3)_c$  & $SU(2)_L$ & $U(1)_Y$ & $U(1)_X$  \\ 
\hline\hline
$q_{L_i}$ & {\bf 3 }    &  {\bf 2}         & $ 1/6$       & $(1/6) x_{H} + (1/3) x_{\Phi}$    \\
$u_{R_i}$ & {\bf 3 }    &  {\bf 1}         & $ 2/3$       & $(2/3) x_{H} + (1/3) x_{\Phi}$  \\
$d_{R_i}$ & {\bf 3 }    &  {\bf 1}         & $-1/3$       & $-(1/3) x_{H} + (1/3) x_{\Phi}$  \\
\hline
$\ell_{L_i}$ & {\bf 1 }    &  {\bf 2}         & $-1/2$       & $(-1/2) x_{H} - x_{\Phi}$  \\
$e_{R_i}$    & {\bf 1 }    &  {\bf 1}         & $-1$                   & $-x_{H} - x_{\Phi}$  \\
\hline
$H$            & {\bf 1 }    &  {\bf 2}         & $- 1/2$       & $(-1/2) x_{H}$  \\  
\hline
$N_{R_i}$    & {\bf 1 }    &  {\bf 1}         &$0$                    & $- x_{\Phi}$   \\

$\Phi$            & {\bf 1 }       &  {\bf 1}       &$ 0$                  & $ + 2x_{\Phi}$  \\ 
\hline\hline
\end{tabular}
\end{center}
\caption{
Particle content of  the minimal U(1)$_X$ model. In addition to the SM particle content, three RHNs ($N_{R_i}$) and 
  a SM-singlet Higgs field ($\Phi$) are introduced. Here $i=1, 2, 3$ is the family index, and $x_H$, $x_\Phi$ are real parameters. Since the U(1)$_X$ gauge coupling is a free parameter in the model, 
  we fix $x_\Phi=1$ without loss of generality. 
}
\label{tab1}
\end{table}  

We first consider the minimal $U(1)_X$ model whose particle content is listed in Table~\ref{tab1}.   
In addition to the SM particle content, three RHNs ($N_{R_i}$) are introduced 
  to cancel the gauge and the mixed gauge-gravitational anomalies.  
A new Higgs field ($\Phi$), which is singlet under the SM gauge group, is 
  also introduced to break the $U(1)_X$ gauge symmetry 
  by its vacuum expectation value (VEV).  
Since the $U(1)_X$ gauge coupling is a free parameter in the model, 
   we can fix $x_\Phi=1$ without loss of generality, and 
   the U(1)$_X$ charge of a particle is thus fixed by the $x_H$ value.  
Note that in the limit of $x_{H} \to 0$ the minimal $U(1)_X$ model is identical to the minimal $B-L$ model.

The Yukawa Lagrangian of the SM is extended to include
\bea
-\mathcal{L} _{Y}\ \supset \ \sum_{i,j=1}^{3} Y_{D}^{ij} \overline{\ell_{L_i}} H N_{R_j}
  +\frac{1}{2} \sum_{i,j=1}^{3} Y_{N}^{ij} \overline{N_{R_i}^{C}} \Phi N_{R_j}+ \rm{H. c.} \, , 
\label{U1XYukawa}
\eea 
where $C$ denotes taking charge-conjugation, and 
  the first and second terms on the right-hand side are the Dirac and Majorana Yukawa couplings, respectively. 
In order to break the electroweak and the $U(1)_X$ gauge symmetries,  
  we assume a suitable Higgs potential for $H$ and $\Phi$ to develop their VEVs 
  \begin{align}
  \langle H \rangle \ = \ \frac{1}{\sqrt{2}}\begin{pmatrix} v\\0 
  \end{pmatrix} \, , \quad {\rm and}\quad 
 \langle \Phi \rangle \ =\  \frac{v_\Phi}{\sqrt{2}} \, ,
  \end{align}
  respectively at the potential minimum (with $v\simeq 246$ GeV and $v_\Phi$ hitherto a free parameter).  After the symmetry breaking, 
  the mass of the $U(1)_X$ gauge boson ($Z^\prime$ boson),  the Majorana masses for the RHNs and the neutrino Dirac masses are generated as follows:
\bea
 m_{Z^\prime} \ & = & \ g_X \sqrt{4 v_\Phi^2+  \frac{1}{4}x_H^2 v^2} \ \simeq \ 2 g_X v_\Phi ,  \\
   m_{N_i} \ & = & \ \frac{Y^i_{N}}{\sqrt{2}} v_\Phi, \label{mNI} \\
   m_{D}^{ij} \ & = & \ \frac{Y_{D}^{ij}}{\sqrt{2}} v, \label{mDI}
\label{masses}   
  \eea   
where $g_X$ is the $U(1)_X$ gauge coupling. Here we have used the LEP~\cite{LEP:2003aa}, Tevatron~\cite{Carena:2004xs} and LHC~\cite{Amrith:2018yfb} constraints which generically imply $m_{Z'}/g_X\gtrsim 6.9$ TeV at 95\% CL (for the $B-L$ case) to assume $v_\Phi^2 \gg {v}^2$. Also, without loss of generality, we have set our basis in which $Y_N$ is diagonal.
With the generation of the Dirac and Majorana masses, type-I seesaw mechanism can be used to account for tiny Majorana masses of the light neutrino mass eigenstates (see Section~\ref{Sec4} for more details).

\subsection{Case-II: Alternative $U(1)_X$ Model}
\begin{table}[t]
\begin{center}
\begin{tabular}{c|c|c|c|c|c}
\hline\hline
      &  $SU(3)_c$  & $SU(2)_L$ & $U(1)_Y$ & $U(1)_{X}$ \\ 
\hline
$q_{L_i}$ & {\bf 3 }    &  {\bf 2}         & $ 1/6$       &  $ (1/6) x_{H} + (1/3)$ \\
$u_{R_i}$ & {\bf 3 }    &  {\bf 1}         & $ 2/3$       & $(2/3) x_{H} + (1/3) $ \\
$d_{R_i}$ & {\bf 3 }    &  {\bf 1}         & $-1/3$       & $-(1/3) x_{H} + (1/3) $\\
\hline
$\ell_{L_i}$ & {\bf 1 }    &  {\bf 2}         & $-1/2$       & $(-1/2) x_{H} - 1 $ \\
$e_{R_i}$    & {\bf 1 }    &  {\bf 1}         & $-1$         & $-x_{H} - 1 $ \\
\hline
$H$            & {\bf 1 }    &  {\bf 2}         & $- 1/2$       & $(-1/2) x_{H}$ \\  
\hline
\hline
$N_{R_{1,2}}$    & {\bf 1 }    &  {\bf 1}         &$0$                    & $- 4 $ \\ 
$N_{R_3}$    & {\bf 1 }    &  {\bf 1}         &$0$                           & $+ 5 $   \\
\hline
$H_E$            & {\bf 1 }       &  {\bf 2}       &$ -1/2$                  & $(-1/2) x_{H}+3 $  \\ 
$\Phi_A$            & {\bf 1 }       &  {\bf 1}       &$ 0$                  & $ +8  $  \\ 
$\Phi_B$            & {\bf 1 }       &  {\bf 1}       &$ 0$                  & $ -10 $  \\ 
\hline\hline
\end{tabular}
\end{center}
\caption{
Minimal particle content of the ``alternative" $U(1)_{X}$-extended SM. 
In addition to the SM particle content, three RHNs ($N_{R_i}$) and three  new Higgs fields ($H_E, \Phi_{A}, \Phi_{B}$) are introduced.  Here $i=1,2,3$ stands for the family index and $x_H$ is a real parameter. 
}
\label{tab2}
\end{table}   

The other model we consider is the alternative $U(1)_X$ model, whose  
minimal particle content is listed in Table~\ref{tab2}.\footnote{
Here, we list the scalar content essential for our discussion in this paper. 
With only this scalar particle content, we have Nambu-Goldstone modes more than 
  those eaten by the weak bosons and $Z^\prime$ boson 
  since mixing mass terms for the scalars are forbidden by the gauge symmetry.   
Thus, we need to introduce additional (SM-singlet) scalar fields to eliminate phenomenologically dangerous massless modes. 
Since there are many possibilities for new scalars and it is easy to arrange a suitable Higgs potential, 
  we do not discuss a complete Higgs sector in this paper. 
}
Except for the alternative $U(1)_X$ charge assignment for the RHNs, 
  the fermion particle content is the same as in Table~\ref{tab1}.  
  Note that when we assume the same charge for two RHNs among three RHNs in total, 
  this alternative charge assignment is a unique choice 
  in order to cancel all the anomalies~\cite{Montero:2007cd}.

For generating neutrino masses, we have introduced additional scalar fields:  
  one $SU(2)$ doublet $H_E$ and two SM-singlets $\Phi_{A,B}$. 
The new Higgs doublet $(H_E)$ generates the neutrino Dirac masses, 
  while the SM-singlet scalars generate the Majorana mass terms for $\{N_{R,1},N_{R,2}\}$ and $N_{R,3}$, respectively. The Yukawa Lagrangian of the SM is extended to include 
\bea
-\mathcal{L} _{Y}& \ \supset \ & \sum_{i=1}^{3} \sum_{j=1}^{2} Y_{D}^{ij} \overline{\ell_{L_i}} H_E N_{R_j}+\frac{1}{2} \sum_{k=1}^{2} Y_{N}^{A,k}  \overline{N_{R_k}^{C}}\Phi_A N_{R_k} 
+\frac{1}{2} Y^{B}_{N} \overline{N_{R_3}^{C}} \Phi_B  N_{R_3}+ \rm{H. c.} \, ,
\label{ExoticYukawa}
\eea 
where we have assumed a basis in which $Y_N^A$ is diagonal, without loss of generality. 
We also assume a suitable potential for the Higgs fields $H$, $H_E$, $\Phi_{A}$, and $\Phi_B$ to develop 
  their  respective VEVs: 
\bea
  \langle H \rangle \ = \  \frac{1}{\sqrt 2}\left(  \begin{array}{c}  
    v \\
    0 \end{array}
\right),   \;  \;  \; \; 
\langle H_E \rangle \ = \   \frac{1}{\sqrt{2}} \left(  \begin{array}{c}  
    \tilde{v}\\
    0 \end{array}
\right),  \;  \;  \; \; 
\langle \Phi_A \rangle \ = \  \frac{v_{A}}{\sqrt{2}},  \;  \;  \; \; 
\langle \Phi_B \rangle \ = \  \frac{v_{B}}{\sqrt{2}}, 
\eea   
with the condition, $v^2 + \tilde{v}^2 = (246 \,  {\rm GeV})^2$. 
After the $U(1)_X$ symmetry breaking, the RHNs and the $U(1)_X$ gauge boson ($Z^\prime$) acquire their masses as follows: 
\bea
 m_{N_{1,2}}&\ =\ &\frac{Y_N^{1,2}}{\sqrt{2}} v_A, \label{mN12II} \\
  m_{N_3} & \ = \ & \frac{Y_N^{3}}{\sqrt{2}} v_B, \label{mN3II} \\ 
 m_{Z^\prime} &=& g_X \sqrt{64 v_{A}^2+ 100 v_{B}^2+  \frac{1}{4} x_H^2 v^2 + \left(-\frac{1}{2} x_H +3\right)^2 {\tilde v}^2} \nonumber \\
 &\simeq& g_X \sqrt{64 v_{A}^2+ 100 v_{B}^2} \, .
\label{masses-Alt}   
\eea 
Again, we have used the collider constraints to set $(v_A^2 + v_B^2) \gg (v^2 + \tilde{v}^2)$.  
The neutrino Dirac mass terms are generated by $\langle H_E \rangle$: 
\bea
m_{D}^{ij} \ = \ \frac{Y_{D}^{ij}}{\sqrt{2}} \, \tilde{v} \, , \label{mDII}
\eea
after which the seesaw mechanism is automatically implemented.
Note that because of the particle content, only two RHNs ($N_{R_{1,2}}$)  are involved in the seesaw mechanism 
   (the so-called ``minimal seesaw'' \cite{Smirnov:1993af, King:1999mb, Frampton:2002qc, Ibarra:2003up, Bambhaniya:2016rbb})    
   while the third RHN ($N_{R_3}$) has no direct coupling with the SM particles 
   and hence it is naturally a dark matter (DM) candidate~\cite{Okada:2018tgy}.

\section{RHN production through $Z^\prime$ boson decay at the LHC}
\label{Sec3}

Since the RHNs are singlet under the SM gauge group, 
   they have no SM gauge interactions in the flavor basis. 
However, a mixing between the RHNs and the SM neutrinos is generated 
   through the Dirac Yukawa coupling in the seesaw mechanism. As a result, the RHN mass eigenstates 
   couple to the weak gauge bosons $W,Z$ through this mixing. 
Although in general this mixing can be made sizable even for TeV-scale RHNs, contrary to the naive seesaw expectations, by virtue of special textures of the Dirac and RHN Majorana mass matrices~\cite{Pilaftsis:1991ug, Tommasini:1995ii, Gluza:2002vs, Xing:2009in, Gavela:2009cd, He:2009ua, Adhikari:2010yt, Deppisch:2010fr, Mitra:2011qr, Dev:2013oxa, Chattopadhyay:2017zvs},  it has been shown \cite{Das:2017nvm} that 
   this mixing has an upper bound of ${\cal O}(0.01)$ to satisfy various experimental constraints, 
   such as the neutrino oscillation data, the electroweak precision measurements, neutrinoless double beta decay 
   and the charged lepton flavor violating (LFV) processes. 
Hence, the canonical production cross section of TeV-scale RHNs through either the weak gauge bosons~\cite{Datta:1993nm, Panella:2001wq, Han:2006ip, delAguila:2007qnc, Dev:2013wba, Alva:2014gxa, Das:2015toa, Das:2016hof, Pascoli:2018heg} or the Higgs boson~\cite{Dev:2012zg, Cely:2012bz, Hessler:2014ssa, Das:2017zjc, Das:2017rsu}  at the LHC is expected to be very small within the minimal seesaw. 

However, in the $U(1)_X$ models under consideration, all SM fermions as well as the RHNs have non-zero $U(1)_X$ charges, and therefore, the RHNs can be efficiently produced at colliders,  
  in particular, through the resonant production of $Z^\prime$ boson, if kinematically allowed, and its subsequent decay   into a pair of RHNs. 
In this section,  we evaluate the production cross section of this process at the LHC: $pp\to Z'\to NN$ for a choice of parameters consistent with the LHC constraints~\cite{Sirunyan:2018exx, CMS:2018wsn, ATLAS:2019} from dilepton channel: $pp\to Z'\to \ell\overline{\ell}$. We also make sure that the dijet constraints~\cite{CMS:2018wxx, ATLAS:2019bov} on $Z'$ are also satisfied. For this purpose, it is useful to explicitly write down the $Z'$ boson partial decay widths into a pair of SM chiral fermions ($f_{L,R}$) (neglecting their masses) 
  and a pair of Majorana RHNs, respectively: 
\bea
\Gamma({Z^{\prime}\to {\overline{f_{L(R)}}} f_{L(R)}}) &\ = \ & N_c \; \frac{g_X^2 }{24 \pi} Q_{f_{L(R)}}^2 m_{Z^{\prime}}, \nonumber \\
\Gamma(Z^{\prime}\to N_i N_i)
 &\ = \ & \frac{g_X^2 }{24\pi} Q_{N_i}^2 m_{Z^{\prime}} \left(1-\frac{4 m^2_{N_i}}{{m^2_{Z^\prime}}}\right)^{3/2}, 
\label{Zpwidths}
\eea 
where $N_c = 1 (3)$ is the color factor for a SM lepton (quark), 
  and $Q_{f_{L(R)}}$, $Q_{N_i}$ are the $U(1)_X$ charges of the respective fermions which can be read off from Table~\ref{tab1} (for Case-I) and Table~\ref{tab2} (for Case-II). The total decay width of the $Z'$ is the sum of partial widths to all SM fermions (quarks and leptons, both left- and right-handed) and the RHNs.

\subsection{Case-I}
\label{C1}
Let us first consider the minimal U(1)$_X$ model with $x_\Phi=1$ (see Table~\ref{tab1}). 
In Ref.~\cite{Okada:2016tci}, the constraints from the LHC 2016 data have been used to obtain an upper bound on the  $U(1)_X$ gauge coupling $g_X$ 
  as a function of $m_{Z^\prime}$. 
The update of the constraints from the LHC 2017 data have been shown in 
  Refs.~\cite{Okada:2018ktp} and \cite{Okada:2017dqs} for $x_H=0$ and $x_H=-0.8$, respectively. 
From this, we can see that the $U(1)_X$ gauge coupling is constrained to be $g_X \lesssim 0.1$ 
  for the $Z^\prime$ boson mass of  a few TeV. 
Recently, the ATLAS collaboration has reported the final result of the LHC Run-2 with 139 fb$^{-1}$ integrated luminosity~\cite{ATLAS:2019}, 
  which significantly improves the previous constraints, as we will show below. 

Since $g_X$ is constrained to be small, 
  the total $Z^\prime$ boson decay width is vary narrow,  
  and thus we use the narrow width approximation (NWA) to evaluate the $Z^\prime$ boson 
  production cross section at the LHC Run-2.  
In this approximation, 
\bea 
   \sigma( pp \to Z^\prime) \ = \ 2 \sum_{q, \, \bar{q}} \int d x \int dy \, f_q(x, Q) \,  f_{\bar{q}} (y, Q) \, \hat{\sigma}(\hat{s}), 
 \label{Xsec1}  
\eea
  where$f_q$ ($ f_{\bar{q}}$) is the parton distribution function (PDF) for a quark (anti-quark), 
            $\hat{s}= x y s$ is the invariant mass squared of the colliding quarks with $\sqrt{s}=13$ TeV 
            for the LHC Run-2, and the NWA cross section of the colliding quarks to produce $Z^\prime$ boson is given by 
\bea
  \hat{\sigma}(\hat{s}) \ = \ \frac{4 \pi^2}{3} \frac{\Gamma(Z^\prime \to q \bar{q})}{m_{Z^\prime}} 
   \, \delta (\hat{s}-m_{Z^\prime}^2) .
 \label{Xsec2}  
\eea             
For the PDFs, we employ CTEQ6L \cite{Pumplin:2002vw} with a factorization scale $Q=m_{Z^\prime}$ for simplicity. 
In our calculation, we scale our result by a $k$-factor of $k=0.947$ 
   to match the recent ATLAS analysis \cite{ATLAS:2019}. 
See Ref.~\cite{Okada:2016gsh} for a procedure to obtain a suitable $k$-factor. 
Note that in the NWA, the cross section is proportional to $g_X^2$.

\begin{figure}
\begin{center}
\includegraphics[scale=0.41]{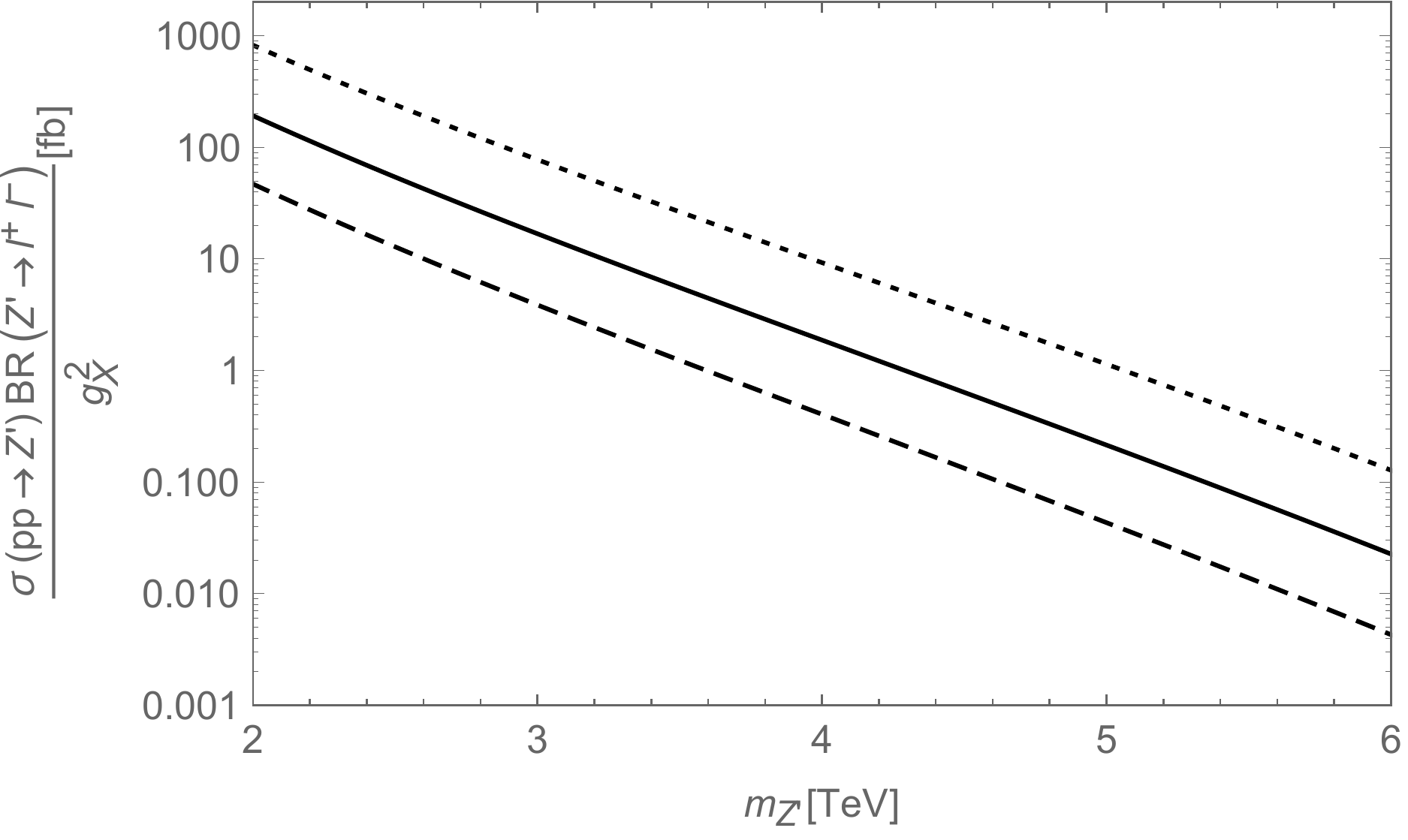} \; \; 
\includegraphics[scale=0.40]{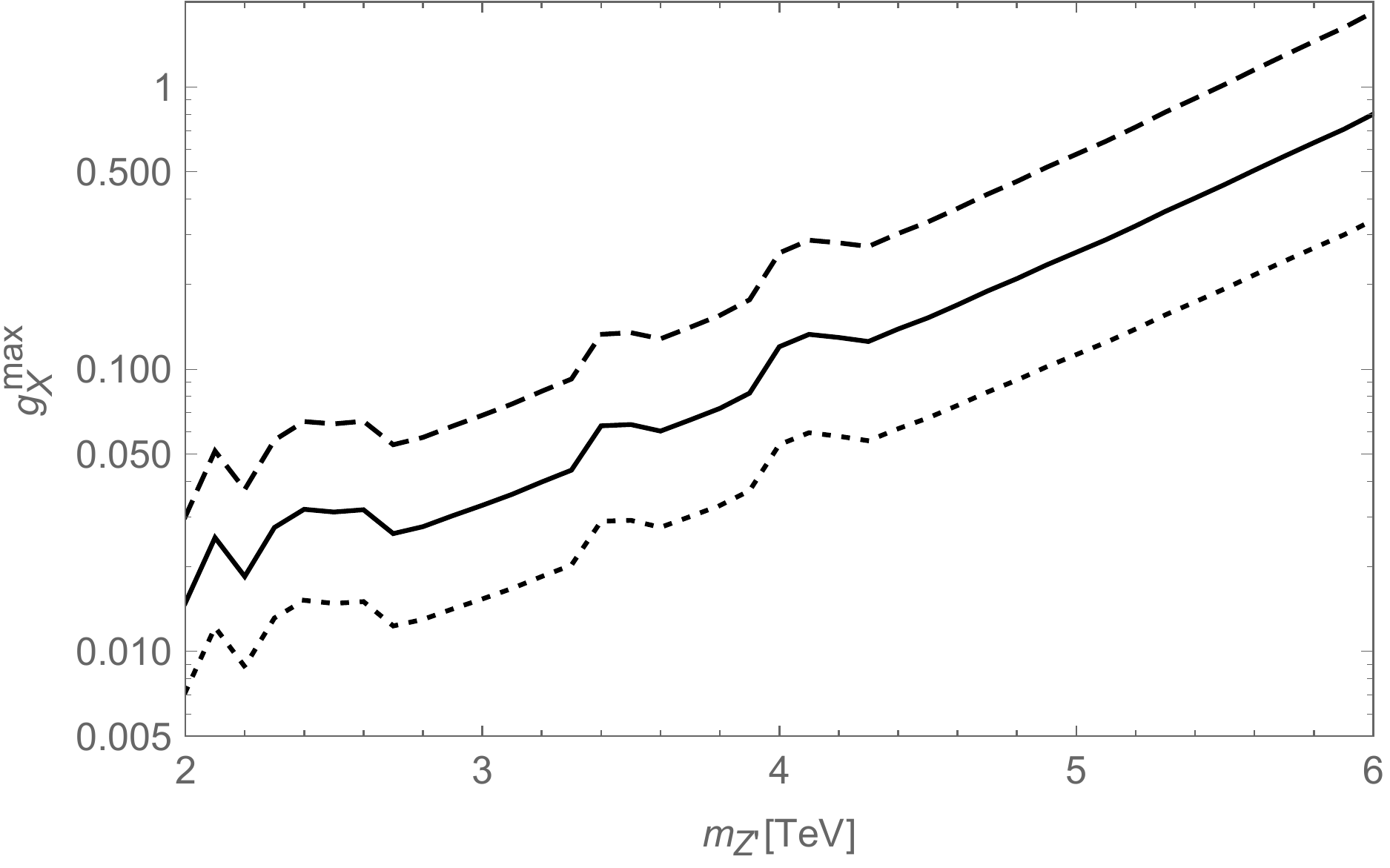}
\caption{
{\it Left:} The dilepton production cross sections (normalized by $g_X^2$) from the $Z^\prime$ boson resonance 
  for $x_H=0$ (solid), $-1.2$ (dashed) and $1$ (dotted) in Case-I.
{\it Right:} The corresponding upper bounds on $g_X$ from the recent ATLAS result. 
}
\label{Zpll}
\end{center}
\end{figure}

\begin{figure}
\begin{center}
\includegraphics[scale=0.41]{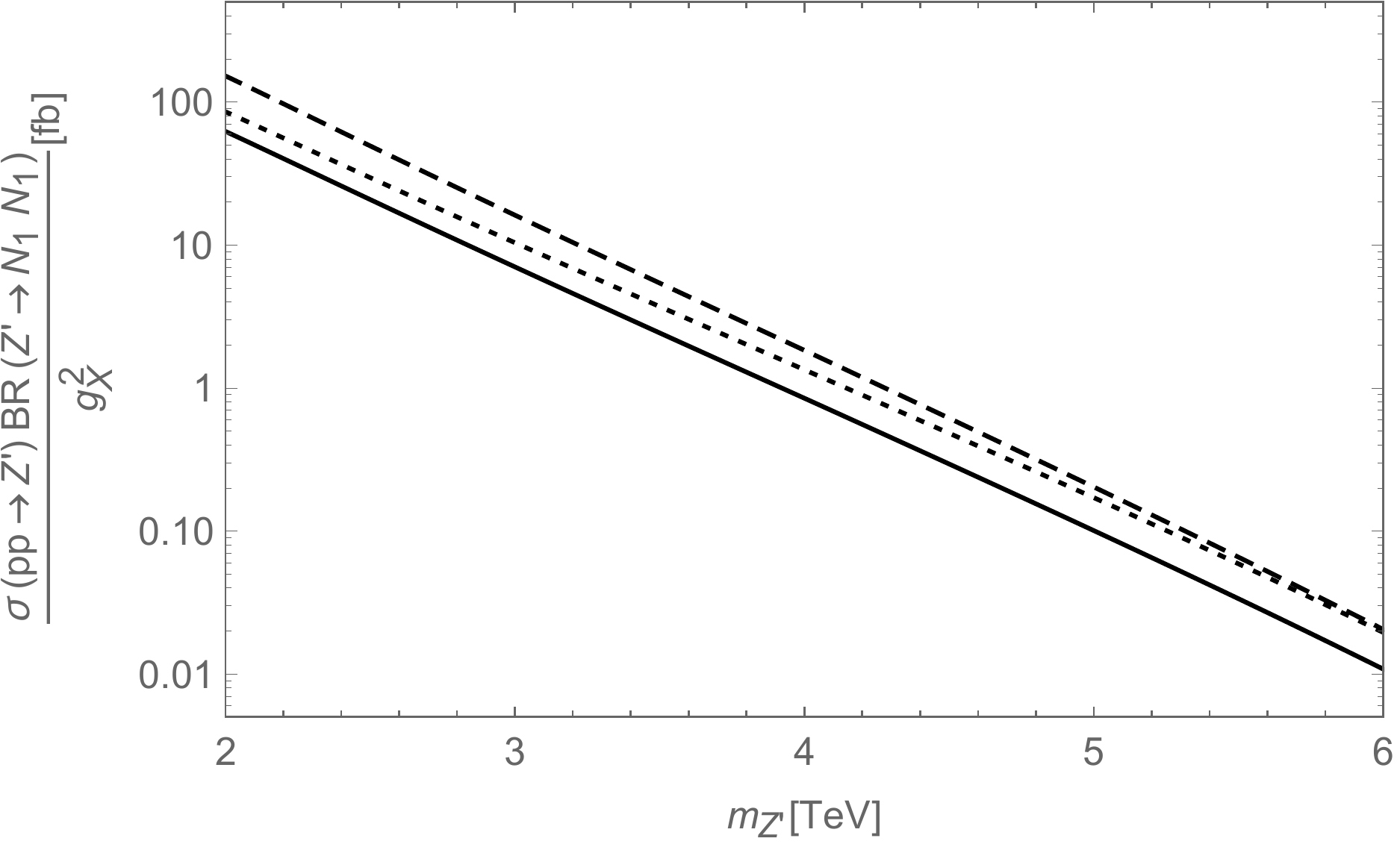}\; \;
\includegraphics[scale=0.41]{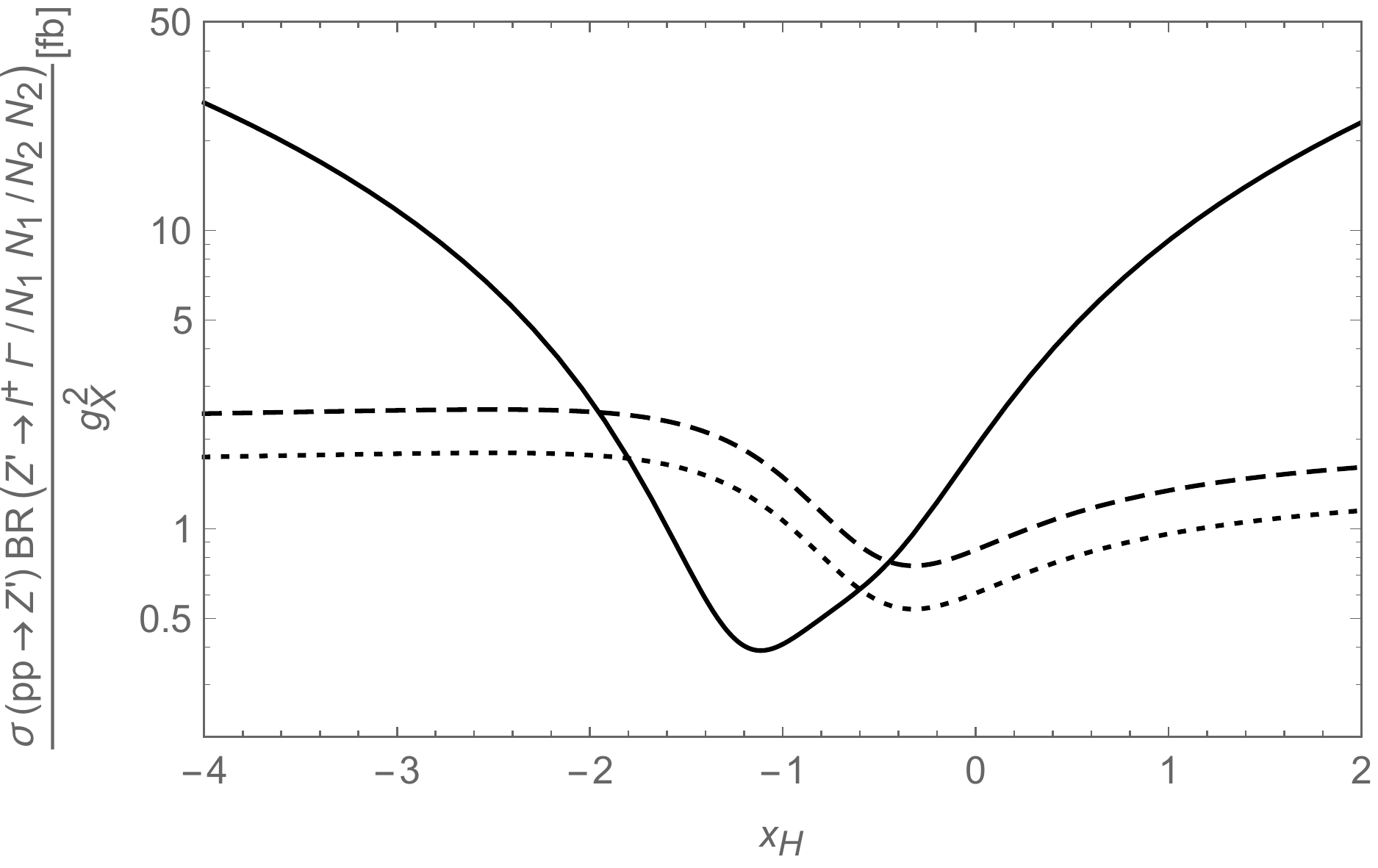} 
\caption{
{\it Left:} The lightest RHN pair-production cross sections (normalized by $g_X^2$) from the $Z^\prime$ boson resonance 
  for $x_H=0$ (solid), $-1.2$ (dashed) and $1$ (dotted) in Case-I.         
{\it Right:} The production cross sections (normalized by $g_X^2$) for dilepton (solid) 
  a pair of $N_1$'s (dashed) and a pair of $N_2$'s (dotted) as a function of $x_H$. Here we have chosen $m_{N_1}=500$ GeV, $m_{N_2}=1$ TeV and $m_{N_3}=2$ TeV.  In the right panel, we have fixed $m_{Z'}=4$ TeV.}
\label{Zp-xH}
\end{center}
\end{figure}

In the left panel of Fig.~\ref{Zpll}, we show the dilepton production cross sections $\sigma(pp\to Z'){\rm BR}(Z'\to \ell^+\ell^-)$
  from the $Z^\prime$ boson resonance 
  for $x_H=0$ (solid), $-1.2$ (dashed) and $1$ (dotted), 
  respectively, as a function of $Z^\prime$ boson mass.  
In this analysis, we have set the RHN mass spectrum as $m_{N_1}=500$ GeV, $m_{N_2}=1$ TeV and $m_{N_3}=2$ TeV for the calculation of the dilepton branching ratio. 
Since the cross section is proportional to $g_X^2$, we have shown the cross section
  normalized by $g_X^2$. 
From this figure, we can read off an upper bound on $g_X$ from the LHC upper limit on the dilepton cross section as a function of the dilepton invariant mass, i.e. $m_{Z^\prime}$ in our case. 
For example, using the recent ATLAS result for the upper bound on 
  $\sigma( pp \to Z^\prime) {\rm BR}(Z^\prime \to \ell^+ \ell^-) \leq$ 0.027 fb for 
  $m_{Z^\prime}=4$ TeV~\cite{ATLAS:2019}, 
   we obtain an upper bound on $g_X \leq0.12$ for $x_H=0$ ($B-L$ limit). 
In the right panel of Fig.~\ref{Zpll}, we show these  upper bounds on $g_X$ 
  for $x_H=0$ (solid), $-1.2$ (dashed) and $1$ (dotted), 
  respectively, as a function of $Z^\prime$ boson mass.

Similarly, in the left panel of Fig.~\ref{Zp-xH},  we show the RHN pair production cross sections 
  from the $Z^\prime$ boson resonance 
  for $x_H=0$ (solid), $-1.2$ (dashed) and $1$ (dotted), 
  respectively, as a function of $Z^\prime$ boson mass.  Here we have chosen $m_{N_1}=500$ GeV, $m_{N_2}=1$ TeV and $m_{N_3}=2$ TeV, as in Fig.~\ref{Zpll}. 
For $m_{Z^\prime}=4$ TeV, we show in the right panel of Fig.~\ref{Zp-xH} the production cross sections 
  for the dilepton (solid), a pair of $N_1$'s (dashed) and a pair of $N_2$'s (dotted)
  as a function of $x_H$. 
We can see that the RHN production cross section is enhanced for $x_H \lesssim -1.5$.   
As has been pointed out in Refs.~\cite{Das:2017flq, Das:2017deo}, 
  the ratio of ${\rm BR}(Z^\prime \to N_i N_i)$ to ${\rm BR}(Z^\prime \to \ell^+ \ell^-)$ 
  is maximized at $x_H=-1.2$. For this choice, the RHN production process from $Z^\prime$ boson resonance is optimized under the severe LHC dilepton constraints.

\subsection{Case-II}
We now repeat the same analysis for the alternative $U(1)_X$ model. 
For simplicity, we assume all extra scalar fields are very heavy and cannot be produced on-shell from $Z^\prime$ boson decay. 
Because of the alternative $U(1)_X$ charge assignment for the RHNs (see Table~\ref{tab2}), 
   the partial decay widths to RHNs in Eq.~(\ref{Zpwidths}) are enhanced. 
As discussed in Ref.~\cite{Okada:2018tgy}, the RHN with $U(1)_X$ charge $-5$ is a natural DM candidate 
   and the observed DM relic density is reproduced with $m_{N_3} \simeq m_{Z^\prime}/2$.  
Taking this possibility into account, we neglect the $Z^\prime$ boson decay process to a pair of $N_{R_3}$. 
\begin{figure}
\begin{center}
\includegraphics[scale=0.41]{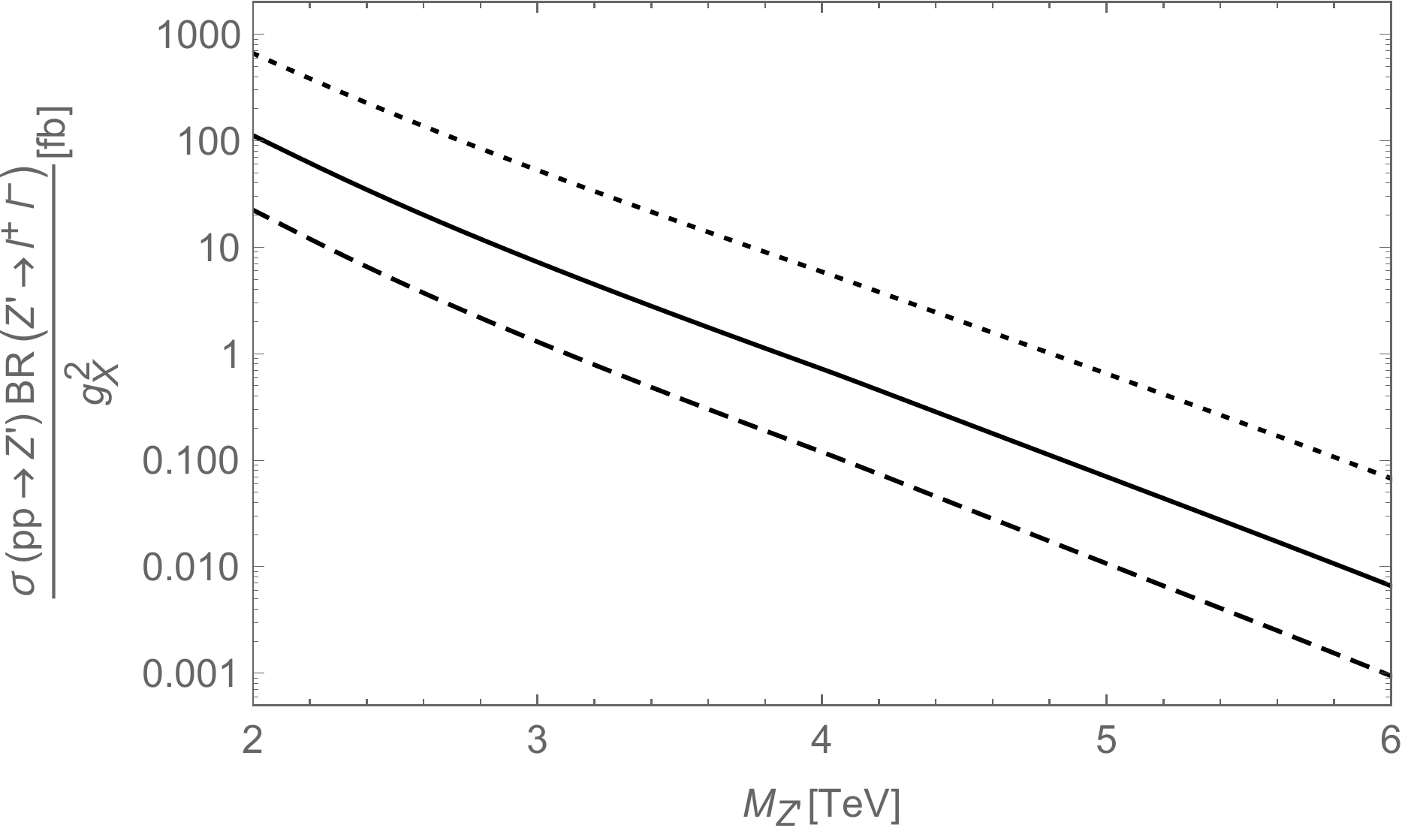} \; \; 
\includegraphics[scale=0.40]{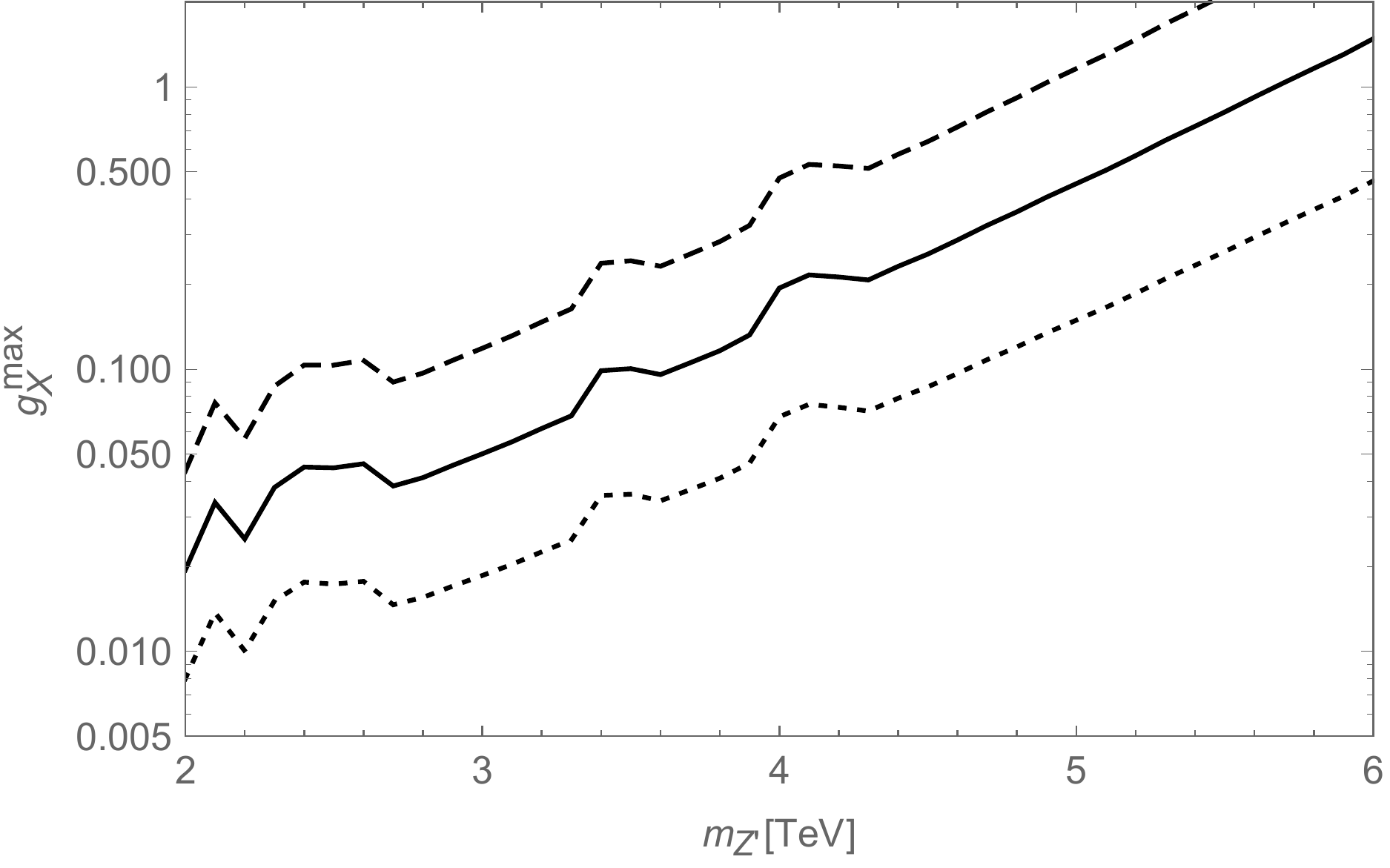}
\caption{
 {\it Left:} The dilepton production cross sections (normalized by $g_X^2$) from the $Z^\prime$ boson resonance 
  for $x_H=0$ (solid), $-1.2$ (dashed) and $1$ (dotted) in Case-II.
{\it Right:} The corresponding upper bounds on $g_X$ from the recent ATLAS result. 
}
\label{Zpll-Alt}
\end{center}
\end{figure}

As in Case-I, we evaluate the $Z^\prime$ boson production cross section by using the NWA. 
In the left panel of Fig.~\ref{Zpll-Alt} (which corresponds to the left panel of Fig.~\ref{Zpll}), 
  we show the dilepton production cross sections from the $Z^\prime$ resonance 
  for $x_H=0$(solid), $x_H=-1.2$ (dashed) and $x_H=1$ (dotted), respectively, 
  as a function of $m_{Z^\prime}$ for $m_{N_1}=500$ GeV and $m_{N_2}=1$TeV.
In the right panel (which corresponds to the right panel of Fig.~\ref{Zpll}), 
  we show the LHC upper bound on $g_X$ derived from recent ATLAS results 
  for $x_H=0$ (solid), $-1.2$ (dashed) and $1$ (dotted), 
  respectively, as a function of $Z^\prime$ boson mass.  
Comparing the results in the left panels of Fig. \ref{Zpll} and \ref{Zpll-Alt},  
  we can see that the dilepton production cross section for the alternative $U(1)_X$ model is 
  smaller than that for the minimal $U(1)_X$ model. 
This is because the partial $Z^\prime$ boson decay widths into the RHNs in the alternative model  
  are much larger than those in the minimal model due to their different charge assignments. 
As a result, the LHC dilepton constraint is milder in this case than that for the minimal $U(1)_X$ model.

\begin{figure}
\begin{center}
\includegraphics[scale=0.41]{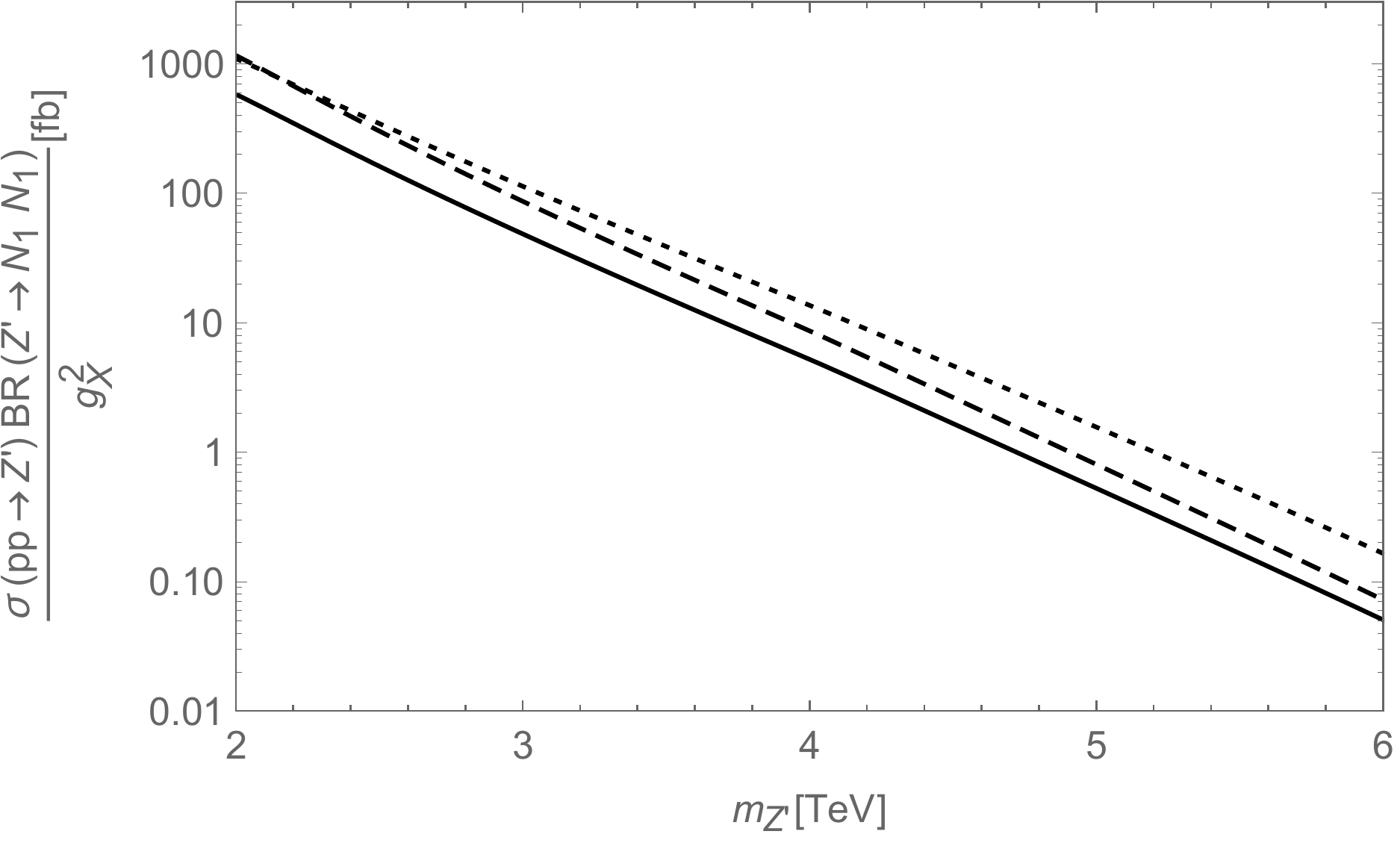}\; \;
\includegraphics[scale=0.41]{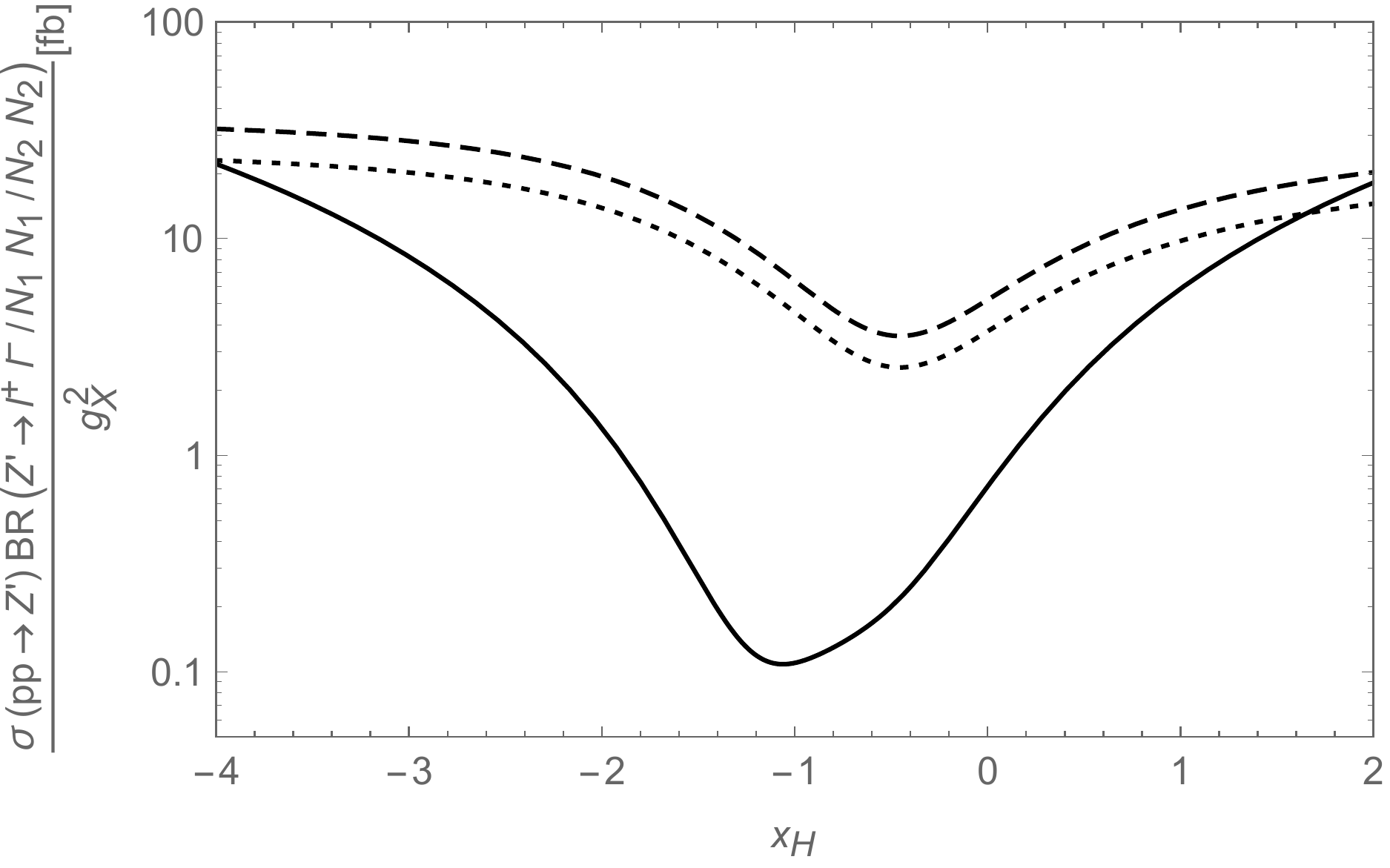} 
\caption{
{\it Left:} The lightest RHN pair-production cross sections (normalized by $g_X^2$) from the $Z^\prime$ boson resonance 
  for $x_H=0$ (solid), $-1.2$ (dashed) and $1$ (dotted) in Case-II.         
{\it Right:} The production cross sections (normalized by $g_X^2$) for dilepton (solid) 
  a pair of $N_1$'s (dashed) and a pair of $N_2$'s (dotted) as a function of $x_H$. Here we have chosen $m_{N_1}=500$ GeV and $m_{N_2}=1$ TeV. In the right panel, we have fixed $m_{Z'}=4$ TeV.
}
\label{Zp-xH-Alt}
\end{center}
\end{figure}

In the left panel of Fig.~\ref{Zp-xH-Alt} (which corresponds to the left panel of Fig.~\ref{Zp-xH}) 
  we show the lightest RHN ($N_1$) pair-production cross section from the $Z^\prime$ boson resonance 
  for $x_H=0$ (solid), $-1.2$ (dashed) and $1$ (dotted), respectively, as a function of $m_{Z^\prime}$.
Comparing with the left panel of Fig.~\ref{Zp-xH}, 
  we can see about an order of magnitude enhancement for the production cross section 
  because of the larger $U(1)_X$ charge of $-4$ for $N_1$. 
For $m_{Z^\prime}=4$ TeV, we show in the right panel of Fig.~\ref{Zp-xH-Alt} (which corresponds to the right panel of Fig.~\ref{Zp-xH}) 
  the production cross sections for the dilepton (solid), a pair of $N_1$'s (dashed) and a pair of $N_2$'s (dotted)
  as a function of $x_H$. 
As in Case-I, the ratio of ${\rm BR}(Z^\prime \to N_i N_i)$ to ${\rm BR}(Z^\prime \to \ell^+ \ell^-)$ 
  is maximized at $x_H=-1.2$ and the RHN production process is optimized under the severe LHC dilepton constraints.

\section{Long-Lived Right-Handed Neutrinos} 
\label{Sec4}

After the breaking of the electroweak and the $U(1)_X$ symmetries, 
  we can write the full neutrino mass matrix as 
\bea
M_{\nu} \ = \ \begin{pmatrix}
0&&m_{D}\\
m_{D}^{T}&&m_{N}
\end{pmatrix} .  
\label{typeInu}
\eea
Without loss of generality, we go to the basis in which the Majorana mass matrix $m_N$ is diagonal, with eigenvalues given in Eqs.~\eqref{mNI} (for Case-I) and \eqref{mN12II}, \eqref{mN3II} (for Case-II). 
 The Dirac mass matrix ($m_D$) elements are given 
  in Eqs.~(\ref{mDI}) (for Case-I) and (\ref{mDII}) (for Case-II).    
Diagonalizing the mass matrix in Eq.~\eqref{typeInu}, we obtain the seesaw formula
 for the light Majorana neutrino mass matrix as 
\bea
m_{\nu}  \ \simeq \ - m_{D} m_{N}^{-1} m_{D}^{T}.
\label{seesawI}
\eea 
 We express the light neutrino flavor eigenstate $(\nu_\alpha)$ in terms of the mass eigenstates 
  of the light $(\nu_i)$ and heavy $(N_i)$ Majorana neutrinos: 
\begin{align}
\nu_\alpha \ = \ \mathcal{N}_{\alpha i} \nu_i+\mathcal{R}_{\alpha i} N_i \, ,
\end{align} 
  where $\mathcal{R} =m_D m_N^{-1}$ characterizes the light-heavy neutrino mixing, $\mathcal{N}=\Big(1-\frac{1}{2}\epsilon\Big) U_{\rm{PMNS}}$ 
    with $\epsilon=\mathcal{R}^\ast\mathcal{R}^T$ the non-unitarity parameter, and $U_{\rm{PMNS}}$ is the $3\times 3$ light neutrino mixing matrix 
   which diagonalizes the light neutrino mass matrix as
\bea
U_{\rm{PMNS}}^T m_\nu U_{\rm{PMNS}} \ = \ {\rm diag}(m_1, m_2, m_3).
\eea
In the presence of $\epsilon$, the mixing matrix $\mathcal{N}$ is not unitary, namely $\mathcal{N}^\dagger\mathcal{N}\neq1$.

In terms of the neutrino mass eigenstates, the charged current $(\rm{CC})$ interaction can be written as 
\bea 
-\mathcal{L}_{\rm{CC}} \ =  \ 
 \frac{g}{\sqrt{2}} W_{\mu}
  \overline{\ell_\alpha} \gamma^{\mu} P_L 
   \left( {\cal N}_{\alpha j} \nu_{j}+ {\cal R}_{\alpha j} N_{j} \right) + \rm{H.c.}, 
\label{CC}
\eea
where $g$ is the $SU(2)_L$ gauge coupling, $\ell_\alpha$ ($\alpha=e, \mu, \tau$) denotes the three generations of SM charged leptons, 
  and $P_L =  \frac{1}{2}(1- \gamma_5)$ is the left-handed projection operator.  
Similarly, the neutral current $(\rm{NC})$ interaction is given by 
\bea 
-\mathcal{L}_{\rm{NC}}& \ = \ & 
 \frac{g}{2 \cos \theta_{w}}  Z_{\mu} 
\Big[ 
  \overline{\nu_{i}} \gamma^{\mu} P_L ({\cal N}^\dagger {\cal N})_{ij} \nu_{{j}}
 +  \overline{N_{{i}}} \gamma^{\mu} P_L ({\cal R}^\dagger {\cal R})_{ij} N_{{j}} \nonumber \\
&& \qquad +\Big\{ 
  \overline{\nu_{{i}}} \gamma^{\mu} P_L ({\cal N}^\dagger  {\cal R})_{ij} N_{{j}} 
  + \rm{H.c.} \Big\} 
\Big] , 
\label{NC}
\eea
where $\theta_{w}$ is the weak mixing angle.  
Through these interactions and the original Dirac Yukawa interactions given in Eqs.~\eqref{U1XYukawa} and \eqref{ExoticYukawa}, 
  the RHNs mainly decay into $N \to \ell W$, $\nu Z$, $\nu h$, 
  where $h$ is the SM Higgs boson. If kinematically allowed, these are two-body decays, followed by the SM decays of the $W,Z,h$. For smaller RHN masses, these decays will be three-body, with off-shell $W,Z,h$.  
Here, we have assumed the $U(1)_X$ Higgs boson(s) is heavier than the RHNs, for simplicity.

In the seesaw model with 2 degenerate RHNs, the upper bound on  ${\cal R}$   
   has been found in Ref.~\cite{Das:2017nvm} as $|\mathcal{R}_{\alpha j}| \lesssim 0.01$ 
   by considering various experimental constraints such as neutrino oscillation data~\cite{Tanabashi:2018oca}, 
   charged LFV phenomena~\cite{Adam, Aubert, OLeary} 
   and electroweak precision measurements~\cite{deBlas:2013gla,delAguila:2008pw, Akhmedov:2013hec}. 
The smallness of the mixing $(\mathcal{R}_{\alpha j})$ between the light and heavy neutrinos 
   implies an RHN mass eigenstate can be long-lived.
If this is the case, a long-lived RHN, once produced at collider experiments through the $Z'$-portal which is unsuppressed by the small mixing, 
  decays into the SM particles after propagating over a measurable distance. 
This displaced vertex phenomenon is a characteristic signature of the production of long-lived particles. 
For RHNs with mass of the TeV-scale scale or smaller,  
  collider searches for the RHNs with displaced vertex provide a promising probe of the seesaw mechanism~\cite{ Antusch:2016ejd}.

Let us now evaluate the lifetime of RHNs in a general parametrization of neutrino mixing. 
We first consider Case-I in which three RHNs are involved in the seesaw mechanism. 
As we will discuss later, the results for Case-II with only two RHNs can be obtained from the results in Case-I in a special limit. 
The elements of the matrix ${\cal R}$ are constrained so as to reproduce the neutrino oscillation data. 
In our analysis, we adopt the following best-fit values for the neutrino oscillation parameters: 
   $\Delta m_{12}^2 = m_2^2-m_1^2 = 7.6 \times 10^{-5}$ eV$^2$, 
   $\Delta m_{23}^2= |m_3^2-m_2^2|=2.4 \times 10^{-3}$ eV$^2$, 
   $\sin^2 2\theta_{12}=0.87$, $\sin^2 2\theta_{23}=1.0$, 
   and  $\sin^{2}2{\theta_{13}}=0.092$, from a recent global fit~\cite{Esteban:2018azc}. 
The $3\times 3$ neutrino mixing matrix is given by 
\bea
U_{\rm{PMNS}} = \begin{pmatrix} c_{12} c_{13}&s_{12}c_{13}&s_{13}e^{i\delta}\\-s_{12}c_{23}-c_{12}s_{23}s_{13}e^{i\delta}&c_{12}c_{23}-s_{12}s_{23}s_{13}e^{i\delta}&s_{23} c_{13}\\ s_{12}c_{23}-c_{12}c_{23}s_{13}e^{i\delta}&-c_{12}s_{23}-s_{12}c_{23}s_{13}e^{i\delta}&c_{23}c_{13} \end{pmatrix} \begin{pmatrix} 1&0&0\\0&e^{i\rho_1}&0\\0&0&e^{i\rho_2}\end{pmatrix}, 
\label{pmns}
\eea
where $c_{ij}=\cos\theta_{ij}$ and $s_{ij}=\sin\theta_{ij}$. 
In our analysis, we set the Dirac $CP$-phase as $\delta=\frac{3\pi}{2}$ as indicated 
 by the recent NO$\nu$A~\cite{Adamson:2016tbq} and T2K~\cite{Abe:2017uxa} data 
 while the Majorana phases $\rho_{1,2}$ are set as free parameters.

We consider both normal hierarchy (NH) where the light neutrino mass eigenvalues are ordered as
  $m_1 < m_2 < m_3$ and inverted hierarchy (IH) 
  where the light neutrino mass eigenvalues are ordered as $m_3 < m_1< m_2$. 
We vary the lightest mass eigenvalue $m_{\rm lightest}$ up to sub-eV scale, to be consistent with the Planck upper limit on the sum of light neutrino masses: $\sum_i m_i < 0.12$ eV~\cite{Aghanim:2018eyx}.   

The seesaw formula allows us to parameterize the mixing angle between the light and heavy neutrinos as~\cite{Casas:2001sr} 
\bea
\mathcal{R}^{\rm{NH/IH}} \ = \ U_{\rm{PMNS}}^{\ast} \sqrt{D^{\rm{NH/ IH}}} \, O \, \sqrt{m_N^{-1}},
\label{gp}
\eea
where $O$ is a general orthogonal matrix: 
\bea
O \ = \ 
\begin{pmatrix}
1&0&0\\
0&\cos x& \sin x\\
0&-\sin x& \cos x
\end{pmatrix}
\begin{pmatrix}
\cos y&0&\sin y\\
0&1& 0\\
-\sin y& 0&\cos y
\end{pmatrix}
\begin{pmatrix}
\cos z&\sin z&0\\
-\sin z&\cos z&0\\
0&0&1
\end{pmatrix}
\label{Omatrix}
\eea
with the angles, $x, y, z$ being complex numbers, and $D_{\rm NH/IH}$ is the light neutrino mass eigenvalue matrix: 
\bea 
  D^{\rm{NH}} \ = \ {\rm diag}
  \left(m_{\rm lightest}, m_2^{\rm{NH}}, m_3^{\rm{NH}} \right),  
\label{DNH}
\eea 
with $m_2^{\rm{NH}}=\sqrt{ \Delta m_{12}^2+m_{\rm lightest}^2}$ 
        and $m_3^{\rm{NH}}=\sqrt{\Delta m_{23}^2 + (m_2^{\rm{NH}})^2}$, 
while the mass eigenvalue matrix for the IH case is 
\bea 
  D^{\rm{IH}} \ = \ {\rm diag}
\left( m_1^{\rm{IH}}, m_2^{\rm{IH}}, m_{\rm lightest} \right) 
\label{DIH}
\eea 
with $m_2^{\rm{IH}}=\sqrt{ \Delta m_{23}^2 + m_{\rm lightest}^2}$ 
        and $m_1^{\rm{IH}}=\sqrt{(m_2^{\rm{IH}})^2- \Delta m_{12}^2}$. In both cases, the RHN mass matrix is defined as  
\bea 
  m_N \ = \ {\rm diag}
\left(m_{N_1},  m_{N_2}, m_{N_3} \right)
\eea 
  with an ordering of  $m_{N_1} \leq  m_{N_2} \leq m_{N_3}$.  Hence, the matrix ${\cal R}$ in Eq.~\eqref{gp} is a function of $\rho_{1,2}$, $m_{\rm lightest}$, $m_{N_i}$ ($i=1,2,3$), 
  and the three complex angles. 
A generalization of the formula of ${\cal R}$ at the one loop level has been studied in Ref.~\cite{Lopez-Pavon:2015cga}, 
  which are however not important for our analysis.

The  two-body partial decay widths of the RHNs are given by~\cite{Atre:2009rg} 
\bea
\Gamma(N_i \rightarrow \ell_\alpha W)^{\rm{NH/IH}} 
 &\ = \ &
  \frac{|\mathcal{R}_{\alpha i}^{\rm{NH/IH}}|^{2}}{16 \pi} 
 \frac{ (m_{N_i}^2 - m_W^2)^2 (m_{N_i}^2+2 m_W^2)}{m_{N_i}^3 v^2} ,
\nonumber \\
\Gamma(N_i \rightarrow \nu^\alpha Z)^{\rm{NH/IH}} 
 &\ = \ &
  \frac{|\mathcal{R}_{\alpha i}^{\rm{NH/IH}}|^{2}}{32 \pi} 
 \frac{ (m_{N_i}^2 - m_Z^2)^2 (m_{N_i}^2+2 m_Z^2)}{m_{N_i}^3 v^2} ,
\nonumber \\
\Gamma(N_i \rightarrow \nu^\alpha h)^{\rm{NH/IH}}
 &\ = \ & \frac{ |\mathcal{R}_{\alpha i}^{\rm{NH/IH}}|^2}{32 \pi}
   \frac{(m_{N_i}^2-m_h^2)^2}{m_{N_i} v^2} \, .
 \label{widths}
\eea 
respectively. 
In the limit of $m_{N_i} \gg m_W, m_Z, m_h$,  
 the ratio among the partial decay widths is found to be 
   $\Gamma(N_i \rightarrow \ell^\alpha W)^{\rm{NH/IH}} : 
      \Gamma(N_i \rightarrow \nu^\alpha Z)^{\rm{NH/IH}} : 
      \Gamma(N_i \rightarrow \nu^\alpha h)^{\rm{NH/IH}} = 2:1:1$. 
This result is consistent with the Goldstone boson equivalence theorem, 
  since the RHN decay originates from the Dirac Yukawa coupling in Eq.~(\ref{U1XYukawa}). 
The total decay width of the RHN $N_i$ is just the sum of the partial widths:  
\bea
\Gamma_{N_i}^{\rm{NH/IH}} \ = \ \sum_{\alpha=e, \mu, \tau} 
    \left[  \Gamma(N_i \rightarrow \ell_\alpha W)^{\rm{NH/IH}} + 
      \Gamma(N_i \rightarrow \nu_\alpha Z)^{\rm{NH/IH}}  +
      \Gamma(N_i \rightarrow \nu_\alpha h)^{\rm{NH/IH}} \right],
\eea       
and the total proper decay length of the RHN $N_i$ is 
\bea
L_i^{\rm{NH/ IH}} \ = \ \frac{1.97\times 10^{-13}}{\Gamma_{N_i}^{\rm{NH/IH}}[{\rm GeV}]}~~ [\rm{mm}] .
\eea

Employing the general parametrization for the neutrino Dirac mass matrix in Eq.~(\ref{gp}), 
  we perform a parameter scan with free parameters, $0 \leq \rho_{1,2} \leq 2 \pi$, $m_{\rm lightest}$, $x$, $y$, and $z$, 
  to evaluate $L_i^{\rm{NH/ IH}}$ while satisfying all the phenomenological constraints listed in Ref.~\cite{Das:2017nvm}.
For concreteness, we fix $m_{N_1}=500$ GeV, $m_{N_2}=1$ TeV and $m_{N_3}=2$ TeV in our analysis. See Ref.~\cite{Das:2017nvm} for a detail of this parameter scan procedure. 
The most stringent lower bound on the decay length of the RHN $N_i$ comes from two experimental constraints.  
The first is from LFV muon decay process of $\mu \to e \gamma$, whose branching ratio must be $\leq 4.2\times 10^{-13}$~\cite{Adam}
  which provides an upper bound on $|\epsilon_{1 2}| < 1.3 \times 10^{-5}$.
The second is from the lower limit on the half-life of neutrino-less double beta decay: $T_{1/2}^{0\nu}(^{76}{\rm Ge})\geq 8\times 10^{25}{\rm yr}$~\cite{Agostini:2018tnm} that translates into an upper limit on the amplitude for the contribution mediated by the RHNs~\cite{Dev:2013vxa, DellOro:2016tmg}:
\bea
  \left| \sum_{j=1}^3   \frac{ {\cal R}_{e j}}{ m_{N_j} [{\rm GeV]}} \right|  \lesssim 7.8 \times 10^{-8} \, .
\eea
\begin{figure}
\begin{center}
\includegraphics[scale=0.245]{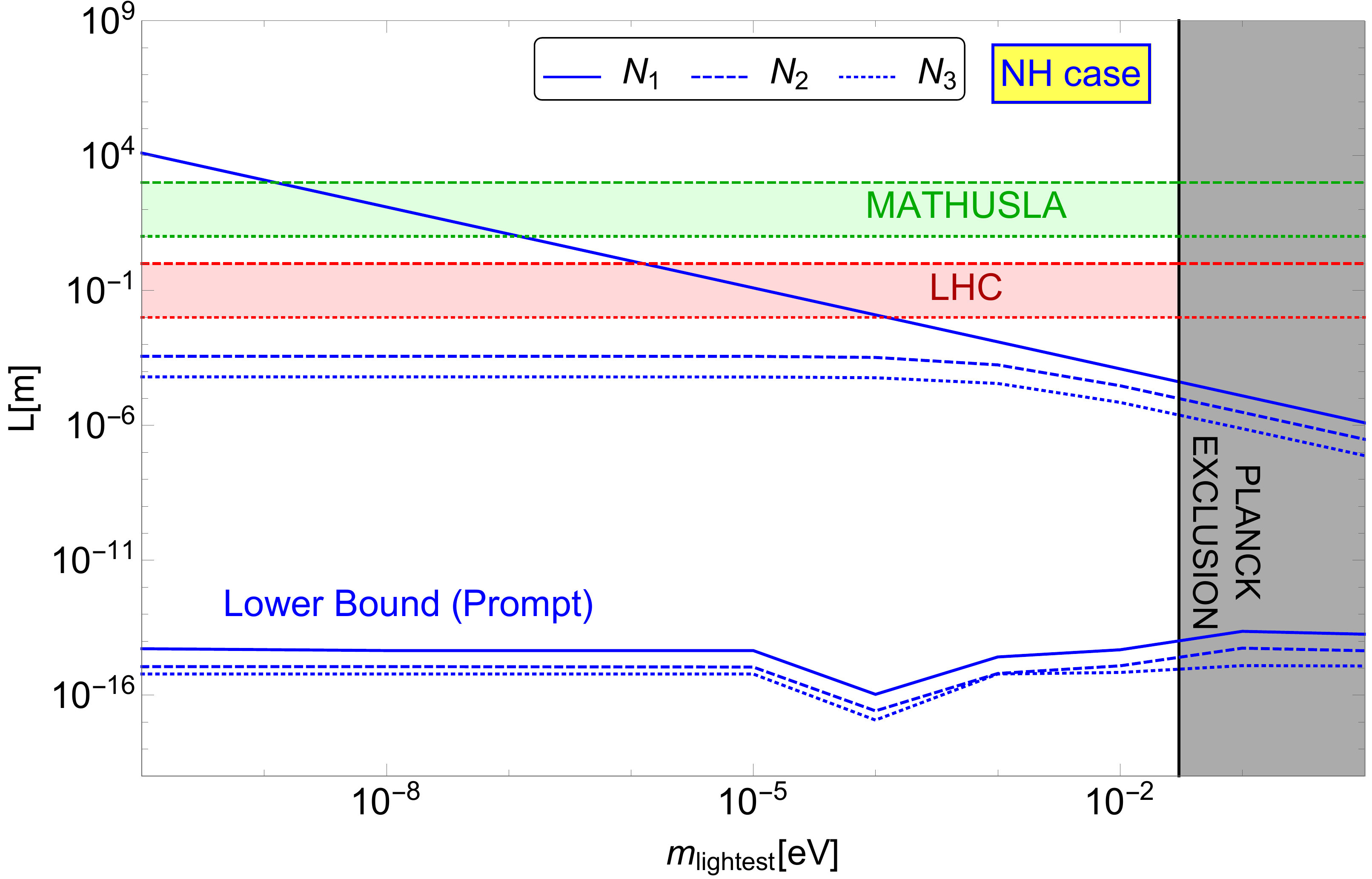} \; 
\includegraphics[scale=0.239]{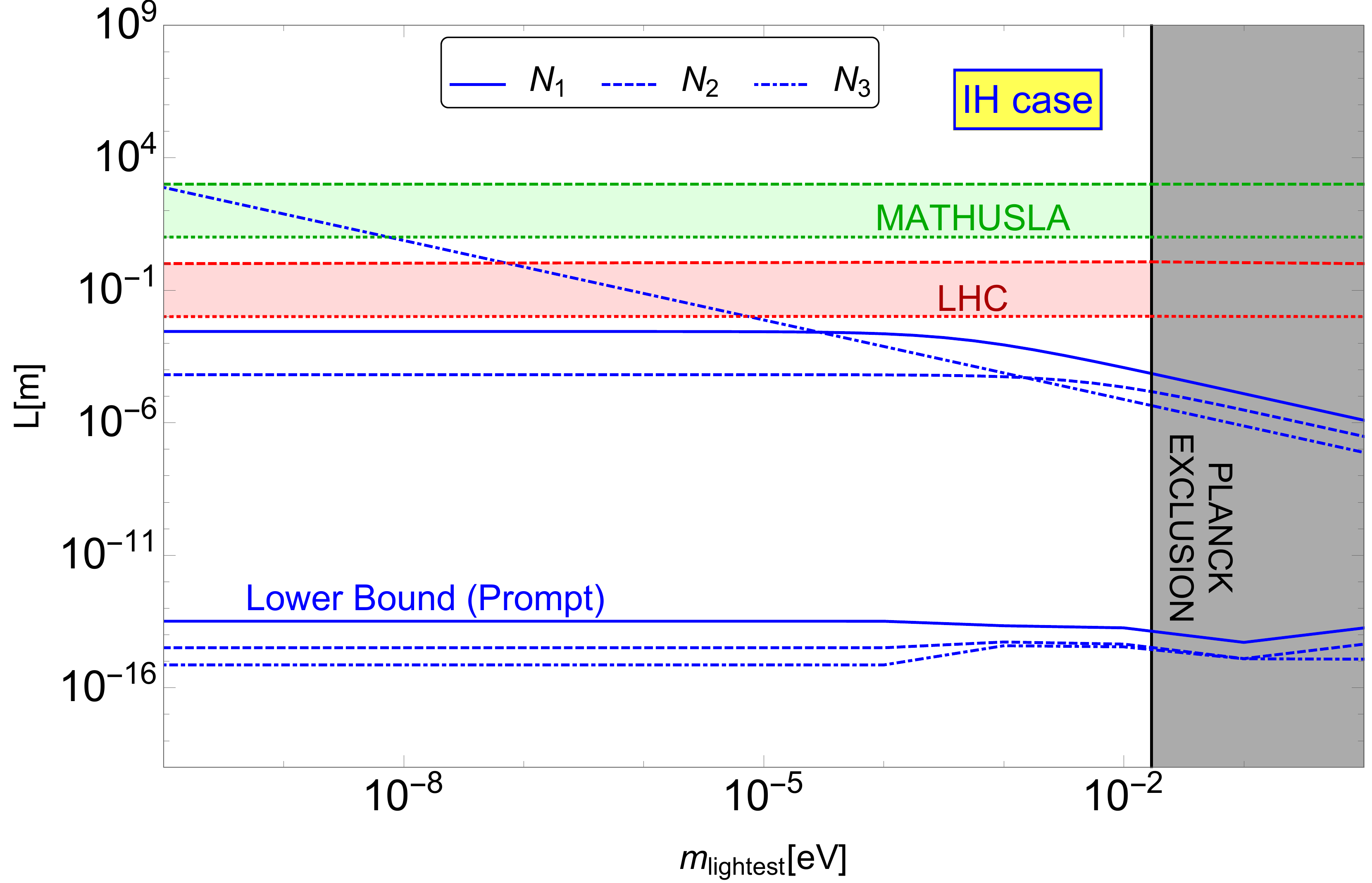}
\caption{
Decay length of RHNs neutrinos as a function of lightest active neutrino mass 
  for the NH (IH) case in the left (right) panel for the three generations of RHNs 
  with $m_{N_1}=500$ GeV, $m_{N_2}=1$ TeV and $m_{N_3}=2$ TeV.  The upper (lower) curves correspond to the maximum (minimum) allowed decay lifetime, taking into account various phenomenological constraints (see text). The horizontal red (green) band indicates the typical range relevant for observable displaced vertex signal at the LHC (MATHUSLA). The vertical shaded region is excluded by Planck upper limit on the sum of neutrino masses. 
}
\label{NH-IH}
\end{center}
\end{figure}

Our results for the upper and lower bounds on $L_i^{\rm{NH/ IH}}$ as a function of the lightest neutrino mass eigenvalue 
  are shown in Fig.~\ref{NH-IH} for the NH (left panel) and IH (right panel) cases in the minimal $U(1)_X$ scenario. We also show as horizontal bands  
 typical decay lengths 
  relevant to the displaced vertex search at the LHC and at MATHUSLA. The vertical shaded region is excluded by the cosmological upper bound on the sum of light neutrino masses $\Sigma_i m_i < 0.12$ eV
  from the Planck 2018 results~\cite{Aghanim:2018eyx}. 
We find that the maximum proper decay length of an RHN can be approximately expressed as 
\bea
L_{{\rm{max}}}^{\rm NH} \ \simeq \ 0.62 \, \left(\frac{0.001 \, \rm{eV}}{m_{\rm{lighest}}} \right) 
  \left(\frac{1\, \rm{TeV}}{m_{N_1}} \right)~~ [\rm{mm}] \, , \label{Ldv-NH} \\ 
L_{{\rm{max}}}^{\rm{IH}} \ \simeq  \ 0.15 \, \left(\frac{0.001 \, \rm{eV}}{m_{\rm{lighest}}} \right) 
  \left(\frac{1\, \rm{TeV}}{m_{N_3}} \right)~~ [\rm{mm}] \, .
\label{Ldv-IH}
\eea 
Very interestingly, $L_{{\rm{max}}}$ is controlled by the lightest neutrino mass eigenvalue $m_{\rm{lighest}}$,  
  and if $m_{\rm{lighest}}$ is small enough, 
  one RHN becomes long-lived even if its mass is of order of 1 TeV. This is contrary to the common lore that RHNs can be long-lived only for the sub-electroweak scale mass range.
We find that for $m_{\rm{lighest}} \lesssim 10^{-5}$ eV ($10^{-8}$ eV), 
  the RHN can be long-lived enough to be explored by the HL-LHC (MATHUSLA).\footnote{A detailed sensitivity study based on the expected number of events, which depends on other details, such as the flavor of the final state lepton and the Lorentz boost factor of the RHN (which depends on the specific production mode, i.e.~the $Z'$ boson mass in our case), is beyond the scope of this paper and is postponed to a future work.}   For a different RHN mass spectrum than that chosen in our illustrative benchmark, the corresponding decay lifetime and the possibility of having a long-lived RHN can be easily obtained from Eqs.~\eqref{Ldv-NH} and \eqref{Ldv-IH}.

In other words, once a displaced vertex signal is observed in future collider experiments, 
  we can measure the decay length and the mass of the RHN 
  from the invariant mass of its decay products.  
Fig.~\ref{NH-IH} indicates that with such measurements 
  we can obtain an upper bound on $m_{\rm lighest}$.  
On the other hand, the remaining two RHNs promptly decay to the SM particles for any value of $m_{\rm{lighest}}$. 
The final state with same-sign dileptons plus jets from the decay of these pair-produced RHNs
   is a ``smoking-gun'' signature of the Majorana nature of the RHNs,  
   with very small SM background. Thus, a combination of the prompt and displaced multilepton searches can be used to probe the RHN sector of the minimal $U(1)_X$ scenario, while at the same time gaining information on the absolute value of the lightest active neutrino mass. This will be complementary to the direct searches for the absolute neutrino mass scale from beta-decay (e.g. KATRIN~\cite{Angrik:2005ep}) and neutrinoless double beta decay~\cite{DellOro:2016tmg}, as well as the cosmological probes of the absolute neutrino mass, such as the CMB-S4~\cite{Abazajian:2016yjj}. 

Now let us consider the lifetime of RHNs for Case-II, in which only two RHNs are involved in the seesaw mechanism. 
In this minimal seesaw, $m_{\rm lightest} =0$ and the light Majorana neutrino mass matrix~\eqref{seesawI} is of rank-2. 
For Case-II, we have several modifications for the formulas in Case-I given above. 
The neutrino mixing matrix has only one Majorana $CP$-phase, so that we set $\rho_2=0$ in Eq.~(\ref{pmns}). 
The mass eigenvalue matrices for the NH and IH cases in Eqs.~(\ref{DNH}) and (\ref{DIH}) are modified to 
\bea 
  D^{\rm{NH}} \ = \ 
\begin{pmatrix}  
  0 & 0 \\  
  m_2^{\rm{NH}} & 0 \\
  0  & m_3^{\rm{NH}} 
\end{pmatrix}  ,  \;  \; \;
\label{DNH-Alt}
  D^{\rm{IH}} \ = \ 
\begin{pmatrix}  
   m_1^{\rm{IH}} & 0 \\
  0  & m_2^{\rm{IH}} \\
   0 & 0 
\end{pmatrix}  .
\label{DIH-Alt}
\eea 
The RHN mass matrix is defined as  
\bea 
  m_N \ = \ {\rm diag}
\left(m_{N_1},  m_{N_2}  \right)
\eea 
  with an ordering of  $m_{N_1} \leq  m_{N_2}$, 
  and the orthogonal matrix in Eq.~(\ref{gp}) is replaced by a general $2\times 2$ orthogonal matrix. 
It is easy to see that the light-heavy neutrino mixing matrix for Case-II 
   can be obtained by the mixing matrix in Case-I with a special parameter fixing. 
The mixing $\mathcal{R}_{\alpha i}^{\rm NH}$ ($i=1, 2$) in Case-II is obtained 
  from $\mathcal{R}_{\alpha j}^{\rm NH}$ ($j=2,3$) in Case-I 
  by fixing $m_{\rm lightest} =0$, $\rho_2=0$, $y=z=0$ in Eq.~(\ref{Omatrix})
  and identifying $m_{N_{2,3}}$ as $m_{N_{1,2}}$.  
Similarly, the mixing $\mathcal{R}_{\alpha i}^{\rm IH}$ ($i=1, 2$) in Case-II is obtained 
  from $\mathcal{R}_{\alpha i}^{\rm IH}$ ($j=1,2$) in Case-I 
  by fixing $m_{\rm lightest} =0$, $\rho_2=0$, $x=y=0$ in Eq.~(\ref{Omatrix}).  
Performing parameter scan for Case-II, we have found that 
  the resultant lifetime of the RHNs is obtained from the results 
  in Case-I in the limit of $m_{\rm lightest} \to 0$. 
Here, we have fixed $m_{N_1}=1$ TeV and $m_{N_2}=2$ TeV for the NH case. 
The resultant lifetime of $N_1$ ($N_2$) can be read off from the lifetime of $N_2$ ($N_3$) 
  in the limit of $m_{\rm lightest} \to 0$ in the left panel of Fig.~\ref{NH-IH}.  
For the IH case, we have fixed $m_{N_1}=500$ GeV and $m_{N_2}=1$ TeV. 
The resultant lifetime of $N_1$ ($N_2$) can be read off from the lifetime of $N_1$ ($N_2$) 
  in the limit of $m_{\rm lightest} \to 0$ in the right panel of Fig.~\ref{NH-IH}.  
For Case-II, we thus find the RHN lifetime is not long enough to be explored by the displaced vertex search
  experiments, while there is still a possibility of observing the RHN pair-production in the prompt multilepton plus jets searches at the HL-LHC experiment,  as pointed out in Ref.~\cite{Das:2017deo}. 
  

\section{Conclusion}
\label{Sec5}

The neutrino mass matrix is a major missing piece in the SM to be supplemented by new physics beyond the SM. 
Type-I seesaw with RHNs is arguably the simplest mechanism to naturally generate the light neutrino mass matrix  
   satisfying the observed mass-squared differences and mixing angles. 
The gauged $U(1)_X$-extended SM is a simple and consistent framework to incorporate three RHNs,  
   whose existence is essential to cancel all the gauge and mixed gauge-gravitational anomalies. 
The Majorana masses for the RHNs responsible for type-I seesaw are generated through the spontaneous $U(1)_X$ gauge symmetry breaking. 
If the $U(1)_X$ symmetry breaking scale is around a few TeV,  the associated $Z^\prime$ boson and the RHNs can be produced on-shell at the LHC directly via the new gauge coupling $g_X$, independent of the light-heavy neutrino mixing. 
We have studied the RHN pair-production at the LHC through resonant production of the $Z^\prime$ boson. 
In order to satisfy the latest LHC constraints from searches for a high-mass narrow resonance 
   with dilepton final states, we have first calculated the dilepton production cross section 
   from the $Z^\prime$ boson resonance and presented our result as a function of $g_X$ and $m_{Z^\prime}$. 
The allowed parameter region consistent with the LHC constraints is easily read off from this result (see Figures~\ref{Zpll} and \ref{Zpll-Alt}).    
Then, we have calculated the RHN pair-production cross section as a function of $g_X$ and $m_{Z^\prime}$,  
   and identified the maximum cross section under the LHC constraints.

We have also investigated the possibility for a (sub) TeV-scale RHN to be long-lived. 
Since the RHNs are singlets under the SM gauge group, 
  their mass eigenstates have couplings with the weak gauge bosons 
  and the Higgs boson only through their mixing with the SM neutrinos. 
Since these couplings are generically suppressed for a TeV-scale seesaw mechanism, 
  some of the RHNs can be long-lived, depending on their mass and the flavor structure of the Dirac Yukawa coupling matrix. 
Employing the general parametrization for the light neutrino mass matrix 
  to reproduce the neutrino oscillation data, we have performed a parameter scan 
  to evaluate the maximum lifetime for an RHN. 
We have found a simple formula for the decay lifetime which is inversely proportional to the lightest neutrino mass eigenvalue [cf.~Eqs.~\eqref{Ldv-NH} and \eqref{Ldv-IH}]. 
In particular, for $m_{\rm lightest}\lesssim 10^{-5}$ eV, one of the RHNs in the minimal $U(1)_X$ model turns out to be long-lived enough for the benchmark point considered here. 
Such a long-lived RHN provides us with a displaced vertex signature which can be explored 
   at the High-Luminosity LHC and other dedicated long-lived particle detectors such as MATHUSLA. 
Once a long-lived RHN is observed in the future, using the correlations found here, we can set an upper bound 
   on the lightest neutrino mass eigenvalue, which will be complementary to the low-energy direct probes of the absolute neutrino mass scale.

\noindent 
{\bf Note Added} \\
While completing this manuscript, we noticed a new paper~\cite{Deppisch:2019kvs} 
   that also investigates the long-lived heavy neutrino production 
   from the $Z^\prime$ boson, but in the $U(1)_{B-L}$ context and focusing on the low RHN mass range.    
   
\section*{Acknowledgments}
This work is supported in part by 
  the Japan Society for the Promotion of Science Postdoctoral Fellowship for Research in Japan (A.D.), 
  and the United States Department of Energy under Grant No.~DE-SC0017987 (B.D) and DE-SC0012447 (N.O.), and also by the US Neutrino Theory Network Program under Grant No. DE-AC02-07CH11359 (B.D.).  B.D. thanks the Fermilab theory group for warm hospitality during the completion of this work.



\begin{thebibliography}{99}
\bibitem{Tanabashi:2018oca} 
  M.~Tanabashi {\it et al.} [Particle Data Group],
  ``Review of Particle Physics,''
  Phys.\ Rev.\ D {\bf 98}, no. 3, 030001 (2018).

   \bibitem{Minkowski:1977sc} 
  P.~Minkowski,
  ``$\mu \to e\gamma$ at a Rate of One Out of $10^{9}$ Muon Decays?,''
  Phys.\ Lett.\  {\bf 67B}, 421 (1977).

\bibitem{Mohapatra:1979ia}
R.~N.~Mohapatra and G.~Senjanovic, ``Neutrino Mass and Spontaneous Parity Violation,'' Phys.~Rev.~Lett.~{\bf 44}, 912 (1980).    



\bibitem{Yanagida:1979as}
T.~Yanagida, ``Horizontal gauge symmetry and masses of neutrinos,''
  Conf.\ Proc.\ C {\bf 7902131}, 95 (1979).
  
\bibitem{GellMann:1980vs}
M.~Gell-Mann, P. Ramond, and R.~Slansky, ``Complex Spinors and Unified Theories,''
  Conf.\ Proc.\ C {\bf 790927}, 315 (1979)
  [arXiv:1306.4669 [hep-th]].
  
  \bibitem{Schechter:1980gr}
J.~Schechter and J.~W.~F.~Valle, ``Neutrino Masses in $SU(2) \otimes U(1)$ Theories,'' Phys.~Rev.~D {\bf 22}, 2227 (1980). 
  


 \bibitem{Drewes:2013gca} 
  M.~Drewes,
  ``The Phenomenology of Right Handed Neutrinos,''
  Int.\ J.\ Mod.\ Phys.\ E {\bf 22}, 1330019 (2013)
  [arXiv:1303.6912 [hep-ph]].  


\bibitem{Keung:1983uu} 
  W.~Y.~Keung and G.~Senjanovic,
  ``Majorana Neutrinos and the Production of the Right-handed Charged Gauge Boson,''
  Phys.\ Rev.\ Lett.\  {\bf 50}, 1427 (1983).
 
 \bibitem{Ibarra:2010xw} 
  A.~Ibarra, E.~Molinaro and S.~T.~Petcov,
  ``TeV Scale See-Saw Mechanisms of Neutrino Mass Generation, the Majorana Nature of the Heavy Singlet Neutrinos and $(\beta\beta)_{0\nu}$-Decay,''
  JHEP {\bf 1009}, 108 (2010)
  [arXiv:1007.2378 [hep-ph]].
  
  \bibitem{Fernandez-Martinez:2016lgt} 
  E.~Fernandez-Martinez, J.~Hernandez-Garcia and J.~Lopez-Pavon,
  ``Global constraints on heavy neutrino mixing,''
  JHEP {\bf 1608}, 033 (2016)
  [arXiv:1605.08774 [hep-ph]].


\bibitem{Das:2017nvm} 
  A.~Das and N.~Okada,
  ``Bounds on heavy Majorana neutrinos in type-I seesaw and implications for collider searches,''
  Phys.\ Lett.\ B {\bf 774}, 32 (2017)
  [arXiv:1702.04668 [hep-ph]].
  
   \bibitem{Datta:1993nm} 
  A.~Datta, M.~Guchait and A.~Pilaftsis,
  ``Probing lepton number violation via majorana neutrinos at hadron supercolliders,''
  Phys.\ Rev.\ D {\bf 50}, 3195 (1994)
  [hep-ph/9311257].
  
  \bibitem{Panella:2001wq} 
  O.~Panella, M.~Cannoni, C.~Carimalo and Y.~N.~Srivastava,
  ``Signals of heavy Majorana neutrinos at hadron colliders,''
  Phys.\ Rev.\ D {\bf 65}, 035005 (2002)
  [hep-ph/0107308].
  
  \bibitem{Han:2006ip} 
  T.~Han and B.~Zhang,
  ``Signatures for Majorana neutrinos at hadron colliders,''
  Phys.\ Rev.\ Lett.\  {\bf 97}, 171804 (2006)
  [hep-ph/0604064].
  
  \bibitem{delAguila:2007qnc} 
  F.~del Aguila, J.~A.~Aguilar-Saavedra and R.~Pittau,
  ``Heavy neutrino signals at large hadron colliders,''
  JHEP {\bf 0710}, 047 (2007)
  [hep-ph/0703261].
  
  \bibitem{Dev:2013wba} 
  P.~S.~B.~Dev, A.~Pilaftsis and U.~k.~Yang,
  ``New Production Mechanism for Heavy Neutrinos at the LHC,''
  Phys.\ Rev.\ Lett.\  {\bf 112}, no. 8, 081801 (2014)
  [arXiv:1308.2209 [hep-ph]].
  
  \bibitem{Alva:2014gxa} 
  D.~Alva, T.~Han and R.~Ruiz,
  ``Heavy Majorana neutrinos from $W\gamma$ fusion at hadron colliders,''
  JHEP {\bf 1502}, 072 (2015)
  [arXiv:1411.7305 [hep-ph]].
  
  \bibitem{Das:2015toa} 
  A.~Das and N.~Okada,
  ``Improved bounds on the heavy neutrino productions at the LHC,''
  Phys.\ Rev.\ D {\bf 93}, no. 3, 033003 (2016)
  [arXiv:1510.04790 [hep-ph]].
  
  
\bibitem{Das:2016hof} 
  A.~Das, P.~Konar and S.~Majhi,
  ``Production of Heavy neutrino in next-to-leading order QCD at the LHC and beyond,''
  JHEP {\bf 1606}, 019 (2016)
  [arXiv:1604.00608 [hep-ph]].
  
  \bibitem{Pascoli:2018heg} 
  S.~Pascoli, R.~Ruiz and C.~Weiland,
  ``Heavy Neutrinos with Dynamic Jet Vetoes: Multilepton Searches at $\sqrt{s} = 14,~27,$ and $100$ TeV,''
  arXiv:1812.08750 [hep-ph].

  
   \bibitem{delAguila:2008cj} 
  F.~del Aguila and J.~A.~Aguilar-Saavedra,
  ``Distinguishing seesaw models at LHC with multi-lepton signals,''
  Nucl.\ Phys.\ B {\bf 813}, 22 (2009)
  [arXiv:0808.2468 [hep-ph]].  
  

  
  \bibitem{Atre:2009rg} 
  A.~Atre, T.~Han, S.~Pascoli and B.~Zhang,
  ``The Search for Heavy Majorana Neutrinos,''
  JHEP {\bf 0905}, 030 (2009)
  [arXiv:0901.3589 [hep-ph]].
  
  
    \bibitem{Deppisch:2015qwa} 
  F.~F.~Deppisch, P.~S.~B.~Dev and A.~Pilaftsis,
  ``Neutrinos and Collider Physics,''
  New J.\ Phys.\  {\bf 17}, no. 7, 075019 (2015)
  [arXiv:1502.06541 [hep-ph]].  
  
   \bibitem{Antusch:2016ejd} 
  S.~Antusch, E.~Cazzato and O.~Fischer,
  ``Sterile neutrino searches at future $e^-e^+$, $pp$, and $e^-p$ colliders,''
  Int.\ J.\ Mod.\ Phys.\ A {\bf 32}, no. 14, 1750078 (2017)
  [arXiv:1612.02728 [hep-ph]].

  \bibitem{Cai:2017mow} 
  Y.~Cai, T.~Han, T.~Li and R.~Ruiz,
  ``Lepton Number Violation: Seesaw Models and Their Collider Tests,''
  Front.\ in Phys.\  {\bf 6}, 40 (2018)
  [arXiv:1711.02180 [hep-ph]].  
  
  \bibitem{Das:2018hph} 
  A.~Das,
  ``Searching for the minimal Seesaw models at the LHC and beyond,''
  Adv.\ High Energy Phys.\  {\bf 2018}, 9785318 (2018)
  [arXiv:1803.10940 [hep-ph]].  
  
  \bibitem{Aad:2015xaa} 
  G.~Aad {\it et al.} [ATLAS Collaboration],
  ``Search for heavy Majorana neutrinos with the ATLAS detector in pp collisions at $ \sqrt{s}=8 $ TeV,''
  JHEP {\bf 1507}, 162 (2015)
  [arXiv:1506.06020 [hep-ex]].
  
  \bibitem{Sirunyan:2018xiv} 
  A.~M.~Sirunyan {\it et al.} [CMS Collaboration],
  ``Search for heavy Majorana neutrinos in same-sign dilepton channels in proton-proton collisions at $ \sqrt{s}=13 $ TeV,''
  JHEP {\bf 1901}, 122 (2019)
  [arXiv:1806.10905 [hep-ex]].



\bibitem{Mohapatra:1980qe} 
  R.~N.~Mohapatra and R.~E.~Marshak,
  ``Local $B-L$ Symmetry of Electroweak Interactions, Majorana Neutrinos and Neutron Oscillations,''
  Phys.\ Rev.\ Lett.\  {\bf 44}, 1316 (1980)
  Erratum: [Phys.\ Rev.\ Lett.\  {\bf 44}, 1643 (1980)].
  
 

\bibitem{Marshak:1979fm} 
  R.~E.~Marshak and R.~N.~Mohapatra,
  ``Quark-Lepton Symmetry and $B-L$ as the $U(1)$ Generator of the Electroweak Symmetry Group,''
  Phys.\ Lett.\  {\bf 91B}, 222 (1980).
  

\bibitem{Wetterich:1981bx} 
  C.~Wetterich,
  ``Neutrino Masses and the Scale of $B-L$ Violation,''
  Nucl.\ Phys.\ B {\bf 187}, 343 (1981).
  
  
\bibitem{Masiero:1982fi} 
  A.~Masiero, J.~F.~Nieves and T.~Yanagida,
  ``$B-L$ Violating Proton Decay and Late Cosmological Baryon Production,''
  Phys.\ Lett.\  {\bf 116B}, 11 (1982).
  
  
\bibitem{Buchmuller:1991ce} 
  W.~Buchmuller, C.~Greub and P.~Minkowski,
  ``Neutrino masses, neutral vector bosons and the scale of $B-L$ breaking,''
  Phys.\ Lett.\ B {\bf 267}, 395 (1991).
  
  
  
  \bibitem{Basso:2008iv} 
  L.~Basso, A.~Belyaev, S.~Moretti and C.~H.~Shepherd-Themistocleous,
  ``Phenomenology of the minimal $B-L$ extension of the Standard model: $Z'$ and neutrinos,''
  Phys.\ Rev.\ D {\bf 80}, 055030 (2009)
  [arXiv:0812.4313 [hep-ph]].
  
  \bibitem{Deppisch:2013cya} 
  F.~F.~Deppisch, N.~Desai and J.~W.~F.~Valle,
  ``Is charged lepton flavor violation a high energy phenomenon?,''
  Phys.\ Rev.\ D {\bf 89}, no. 5, 051302 (2014)
  [arXiv:1308.6789 [hep-ph]].
  
  \bibitem{Kang:2015uoc} 
  Z.~Kang, P.~Ko and J.~Li,
  ``New Avenues to Heavy Right-handed Neutrinos with Pair Production at Hadronic Colliders,''
  Phys.\ Rev.\ D {\bf 93}, no. 7, 075037 (2016)
  [arXiv:1512.08373 [hep-ph]].


\bibitem{Cox:2017eme} 
  P.~Cox, C.~Han and T.~T.~Yanagida,
  ``LHC Search for Right-handed Neutrinos in $Z^\prime$ Models,''
  JHEP {\bf 1801}, 037 (2018)
  [arXiv:1707.04532 [hep-ph]].
  

\bibitem{Accomando:2017qcs} 
  E.~Accomando, L.~Delle Rose, S.~Moretti, E.~Olaiya and C.~H.~Shepherd-Themistocleous,
  ``Extra Higgs boson and $Z^\prime$ as portals to signatures of heavy neutrinos at the LHC,''
  JHEP {\bf 1802}, 109 (2018)
  [arXiv:1708.03650 [hep-ph]].  
 
   


\bibitem{Sirunyan:2018exx} 
  A.~M.~Sirunyan {\it et al.} [CMS Collaboration],
  ``Search for high-mass resonances in dilepton final states in proton-proton collisions at $\sqrt{s}=$ 13 TeV,''
  JHEP {\bf 1806}, 120 (2018)
  [arXiv:1803.06292 [hep-ex]].
  
  \bibitem{CMS:2018wsn} 
  CMS Collaboration, 
  ``Search for high mass resonances in dielectron final state,''
  CMS-PAS-EXO-18-006.
  
\bibitem{ATLAS:2019}
  G.~Aad {\it et al.} [ATLAS Collaboration],
  ``Search for high-mass dilepton resonances using 139 fb$^{-1}$ of $pp$ collision data collected at $\sqrt{s}=$13 TeV with the ATLAS detector,''
  arXiv:1903.06248 [hep-ex].
  
  \bibitem{Appelquist:2002mw} 
  T.~Appelquist, B.~A.~Dobrescu and A.~R.~Hopper,
  ``Nonexotic Neutral Gauge Bosons,''
  Phys.\ Rev.\ D {\bf 68}, 035012 (2003)
  [hep-ph/0212073].  

  
  \bibitem{Das:2017flq} 
  A.~Das, N.~Okada and D.~Raut,
  ``Enhanced pair production of heavy Majorana neutrinos at the LHC,''
  Phys.\ Rev.\ D {\bf 97}, no. 11, 115023 (2018)
  [arXiv:1710.03377 [hep-ph]].


\bibitem{Das:2017deo} 
  A.~Das, N.~Okada and D.~Raut,
  ``Heavy Majorana neutrino pair productions at the LHC in minimal U(1) extended Standard Model,''
  Eur.\ Phys.\ J.\ C {\bf 78}, no. 9, 696 (2018)
  [arXiv:1711.09896 [hep-ph]].
  
  
   
    \bibitem{Helo:2013esa} 
  J.~C.~Helo, M.~Hirsch and S.~Kovalenko,
  ``Heavy neutrino searches at the LHC with displaced vertices,''
  Phys.\ Rev.\ D {\bf 89}, 073005 (2014)
  Erratum: [Phys.\ Rev.\ D {\bf 93}, no. 9, 099902 (2016)]
  [arXiv:1312.2900 [hep-ph]].
  
  \bibitem{Izaguirre:2015pga} 
  E.~Izaguirre and B.~Shuve,
  ``Multilepton and Lepton Jet Probes of Sub-Weak-Scale Right-Handed Neutrinos,''
  Phys.\ Rev.\ D {\bf 91}, no. 9, 093010 (2015)
  [arXiv:1504.02470 [hep-ph]].
  
  \bibitem{Alekhin:2015byh} 
  S.~Alekhin {\it et al.},
  ``A facility to Search for Hidden Particles at the CERN SPS: the SHiP physics case,''
  Rept.\ Prog.\ Phys.\  {\bf 79}, no. 12, 124201 (2016)
  [arXiv:1504.04855 [hep-ph]].
  
  \bibitem{Gago:2015vma} 
  A.~M.~Gago, P.~Hernández, J.~Jones-Pérez, M.~Losada and A.~Moreno Briceño,
  ``Probing the Type I Seesaw Mechanism with Displaced Vertices at the LHC,''
  Eur.\ Phys.\ J.\ C {\bf 75}, no. 10, 470 (2015)
  [arXiv:1505.05880 [hep-ph]].
  
   \bibitem{Antusch:2016vyf} 
  S.~Antusch, E.~Cazzato and O.~Fischer,
  ``Displaced vertex searches for sterile neutrinos at future lepton colliders,''
  JHEP {\bf 1612}, 007 (2016)
  [arXiv:1604.02420 [hep-ph]].    
  
  \bibitem{Caputo:2017pit} 
  A.~Caputo, P.~Hernandez, J.~Lopez-Pavon and J.~Salvado,
  ``The seesaw portal in testable models of neutrino masses,''
  JHEP {\bf 1706}, 112 (2017)
  [arXiv:1704.08721 [hep-ph]].
  
   \bibitem{Antusch:2017hhu} 
  S.~Antusch, E.~Cazzato and O.~Fischer,
  ``Sterile neutrino searches via displaced vertices at LHCb,''
  Phys.\ Lett.\ B {\bf 774}, 114 (2017)
  [arXiv:1706.05990 [hep-ph]].  

 \bibitem{Dube:2017jgo} 
  S.~Dube, D.~Gadkari and A.~M.~Thalapillil,
  ``Lepton-Jets and Low-Mass Sterile Neutrinos at Hadron Colliders,''
  Phys.\ Rev.\ D {\bf 96}, no. 5, 055031 (2017)
  [arXiv:1707.00008 [hep-ph]].  


    \bibitem{Cottin:2018kmq} 
  G.~Cottin, J.~C.~Helo and M.~Hirsch,
  ``Searches for light sterile neutrinos with multitrack displaced vertices,''
  Phys.\ Rev.\ D {\bf 97}, no. 5, 055025 (2018)
  [arXiv:1801.02734 [hep-ph]].
  
   \bibitem{Kling:2018wct} 
  F.~Kling and S.~Trojanowski,
  ``Heavy Neutral Leptons at FASER,''
  Phys.\ Rev.\ D {\bf 97}, no. 9, 095016 (2018)
  [arXiv:1801.08947 [hep-ph]].        
  
    \bibitem{Cottin:2018nms} 
  G.~Cottin, J.~C.~Helo and M.~Hirsch,
  ``Displaced vertices as probes of sterile neutrino mixing at the LHC,''
  Phys.\ Rev.\ D {\bf 98}, no. 3, 035012 (2018)
  [arXiv:1806.05191 [hep-ph]].  

  \bibitem{Abada:2018sfh} 
  A.~Abada, N.~Bernal, M.~Losada and X.~Marcano,
  ``Inclusive Displaced Vertex Searches for Heavy Neutral Leptons at the LHC,''
  JHEP {\bf 1901}, 093 (2019)
  [arXiv:1807.10024 [hep-ph]].  
  
  \bibitem{SHiP:2018xqw} 
  C.~Ahdida {\it et al.} [SHiP Collaboration],
  ``Sensitivity of the SHiP experiment to Heavy Neutral Leptons,''
  JHEP {\bf 1904}, 077 (2019)
  [arXiv:1811.00930 [hep-ph]].
  
  \bibitem{Dercks:2018wum} 
  D.~Dercks, H.~K.~Dreiner, M.~Hirsch and Z.~S.~Wang,
  ``Long-Lived Fermions at AL3X,''
  Phys.\ Rev.\ D {\bf 99}, no. 5, 055020 (2019)
  [arXiv:1811.01995 [hep-ph]].
  
   \bibitem{Dib:2019ztn} 
  C.~O.~Dib, C.~S.~Kim and S.~Tapia Araya,
  ``Search of light sterile neutrinos from $W^\pm$ decays,''
  arXiv:1903.04905 [hep-ph].    

  
  \bibitem{Drewes:2019fou} 
  M.~Drewes and J.~Hajer,
  ``Heavy Neutrinos in displaced vertex searches at the LHC and HL-LHC,''
  arXiv:1903.06100 [hep-ph].  
  
\bibitem{Bondarenko:2019tss} 
  K.~Bondarenko, A.~Boyarsky, M.~Ovchynnikov, O.~Ruchayskiy and L.~Shchutska,
  ``Probing new physics with displaced vertices: muon tracker at CMS,''
  arXiv:1903.11918 [hep-ph].
  
  \bibitem{Liu:2019ayx} 
  J.~Liu, Z.~Liu, L.~T.~Wang and X.~P.~Wang,
  ``Seeking for sterile neutrinos with displaced leptons at the LHC,''
  arXiv:1904.01020 [hep-ph].
  
   \bibitem{Accomando:2016rpc} 
  E.~Accomando, L.~Delle Rose, S.~Moretti, E.~Olaiya and C.~H.~Shepherd-Themistocleous,
  ``Novel SM-like Higgs decay into displaced heavy neutrino pairs in $U(1)'$ models,''
  JHEP {\bf 1704}, 081 (2017)
  [arXiv:1612.05977 [hep-ph]].

  
  
  \bibitem{Deppisch:2018eth} 
  F.~F.~Deppisch, W.~Liu and M.~Mitra,
  ``Long-lived Heavy Neutrinos from Higgs Decays,''
  JHEP {\bf 1808}, 181 (2018)
  [arXiv:1804.04075 [hep-ph]]. 
  
  
  \bibitem{Jana:2018rdf} 
  S.~Jana, N.~Okada and D.~Raut,
  ``Displaced vertex signature of type-I seesaw model,''
  Phys.\ Rev.\ D {\bf 98}, no. 3, 035023 (2018)
  [arXiv:1804.06828 [hep-ph]].  
  
  \bibitem{Das:2018tbd} 
  A.~Das, N.~Okada, S.~Okada and D.~Raut,
  ``Probing the seesaw mechanism at the 250 GeV ILC,''
  arXiv:1812.11931 [hep-ph].
  
   \bibitem{Deppisch:2019kvs} 
  F.~F.~Deppisch, S.~Kulkarni and W.~Liu,
  ``Heavy neutrino production via $Z'$ at the lifetime frontier,''
  arXiv:1905.11889 [hep-ph].   
        
  \bibitem{Sirunyan:2018mtv} 
  A.~M.~Sirunyan {\it et al.} [CMS Collaboration],
  ``Search for heavy neutral leptons in events with three charged leptons in proton-proton collisions at $\sqrt{s} =$ 13 TeV,''
  Phys.\ Rev.\ Lett.\  {\bf 120}, no. 22, 221801 (2018)
  [arXiv:1802.02965 [hep-ex]].
  
  \bibitem{Aad:2019kiz} 
  G.~Aad {\it et al.} [ATLAS Collaboration],
  ``Search for heavy neutral leptons in decays of $W$ bosons produced in 13 TeV $pp$ collisions using prompt and displaced signatures with the ATLAS detector,''
  arXiv:1905.09787 [hep-ex].      
 

  
    \bibitem{Alimena:2019zri} 
  J.~Alimena {\it et al.},
  ``Searching for long-lived particles beyond the Standard Model at the Large Hadron Collider,''
  arXiv:1903.04497 [hep-ex].

\bibitem{talk} \url{https://indico.cern.ch/event/681549/contributions/2946915/attachments/1664427/2667666/LLP_HL-LHC_LHCP.pdf}



  
 \bibitem{Gligorov:2017nwh} 
  V.~V.~Gligorov, S.~Knapen, M.~Papucci and D.~J.~Robinson,
  ``Searching for Long-lived Particles: A Compact Detector for Exotics at LHCb,''
  Phys.\ Rev.\ D {\bf 97}, no. 1, 015023 (2018)
  [arXiv:1708.09395 [hep-ph]].
  
     \bibitem{Curtin:2018mvb} 
  D.~Curtin {\it et al.},
  ``Long-Lived Particles at the Energy Frontier: The MATHUSLA Physics Case,''
  arXiv:1806.07396 [hep-ph].
  
   \bibitem{Gligorov:2018vkc} 
  V.~V.~Gligorov, S.~Knapen, B.~Nachman, M.~Papucci and D.~J.~Robinson,
  ``Leveraging the ALICE/L3 cavern for long-lived particle searches,''
  Phys.\ Rev.\ D {\bf 99}, no. 1, 015023 (2019)
  [arXiv:1810.03636 [hep-ph]].


  
  \bibitem{Ariga:2018uku} 
  A.~Ariga {\it et al.} [FASER Collaboration],
  ``FASER’s physics reach for long-lived particles,''
  Phys.\ Rev.\ D {\bf 99}, no. 9, 095011 (2019)
  [arXiv:1811.12522 [hep-ph]].
  
 
  
  \bibitem{Pinfold:2019nqj} 
  J.~L.~Pinfold,
  ``The MoEDAL Experiment at the LHC—A Progress Report,''
  Universe {\bf 5}, no. 2, 47 (2019).


  \bibitem{LEP:2003aa} 
  t.~S.~Electroweak [LEP and ALEPH and DELPHI and L3 and OPAL Collaborations and LEP Electroweak Working Group and SLD Electroweak Group and SLD Heavy Flavor Group],
  ``A Combination of preliminary electroweak measurements and constraints on the standard model,''
  hep-ex/0312023.


\bibitem{Carena:2004xs} 
  M.~Carena, A.~Daleo, B.~A.~Dobrescu and T.~M.~P.~Tait,
  ``$Z^\prime$ gauge bosons at the Tevatron,''
  Phys.\ Rev.\ D {\bf 70}, 093009 (2004)
  [hep-ph/0408098].
  
 \bibitem{Amrith:2018yfb} 
  S.~Amrith, J.~M.~Butterworth, F.~F.~Deppisch, W.~Liu, A.~Varma and D.~Yallup,
  ``LHC constraints on a $B-L$ gauge model using Contur,''
  JHEP {\bf 1905}, 154 (2019)
  [arXiv:1811.11452 [hep-ph]].

   \bibitem{Montero:2007cd} 
  J.~C.~Montero and V.~Pleitez,
  ``Gauging $U(1)$ symmetries and the number of right-handed neutrinos,''
  Phys.\ Lett.\ B {\bf 675}, 64 (2009)
  [arXiv:0706.0473 [hep-ph]]. 

 \bibitem{Smirnov:1993af} 
  A.~Y.~Smirnov,
  ``Seesaw enhancement of lepton mixing,''
  Phys.\ Rev.\ D {\bf 48}, 3264 (1993)
  [hep-ph/9304205].
  
\bibitem{King:1999mb} 
  S.~F.~King,
  ``Large mixing angle MSW and atmospheric neutrinos from single right-handed neutrino dominance and $U(1)$ family symmetry,''
  Nucl.\ Phys.\ B {\bf 576}, 85 (2000)
  [hep-ph/9912492].


\bibitem{Frampton:2002qc} 
  P.~H.~Frampton, S.~L.~Glashow and T.~Yanagida,
  ``Cosmological sign of neutrino CP violation,''
  Phys.\ Lett.\ B {\bf 548}, 119 (2002)
  [hep-ph/0208157].
  
  \bibitem{Ibarra:2003up} 
  A.~Ibarra and G.~G.~Ross,
  ``Neutrino phenomenology: The Case of two right-handed neutrinos,''
  Phys.\ Lett.\ B {\bf 591}, 285 (2004)
  [hep-ph/0312138].
  
  \bibitem{Bambhaniya:2016rbb} 
  G.~Bambhaniya, P.~S.~B.~Dev, S.~Goswami, S.~Khan and W.~Rodejohann,
  ``Naturalness, Vacuum Stability and Leptogenesis in the Minimal Seesaw Model,''
  Phys.\ Rev.\ D {\bf 95}, no. 9, 095016 (2017)
  [arXiv:1611.03827 [hep-ph]].
 
 
\bibitem{Okada:2018tgy} 
  N.~Okada, S.~Okada and D.~Raut,
  ``A natural $Z^\prime$-portal Majorana dark matter in alternative U(1) extended Standard Model,''
  arXiv:1811.11927 [hep-ph]. 
 
 
  \bibitem{Pilaftsis:1991ug} 
  A.~Pilaftsis,
  ``Radiatively induced neutrino masses and large Higgs neutrino couplings in the standard model with Majorana fields,''
  Z.\ Phys.\ C {\bf 55}, 275 (1992)
  [hep-ph/9901206].
 
 \bibitem{Tommasini:1995ii} 
  D.~Tommasini, G.~Barenboim, J.~Bernabeu and C.~Jarlskog,
  ``Nondecoupling of heavy neutrinos and lepton flavor violation,''
  Nucl.\ Phys.\ B {\bf 444}, 451 (1995)
  [hep-ph/9503228].
  
  \bibitem{Gluza:2002vs} 
  J.~Gluza,
  ``On teraelectronvolt Majorana neutrinos,''
  Acta Phys.\ Polon.\ B {\bf 33}, 1735 (2002)
  [hep-ph/0201002].
  
  \bibitem{Xing:2009in} 
  Z.~z.~Xing,
  ``Naturalness and Testability of TeV Seesaw Mechanisms,''
  Prog.\ Theor.\ Phys.\ Suppl.\  {\bf 180}, 112 (2009)
  [arXiv:0905.3903 [hep-ph]].
  
  \bibitem{Gavela:2009cd} 
  M.~B.~Gavela, T.~Hambye, D.~Hernandez and P.~Hernandez,
  ``Minimal Flavour Seesaw Models,''
  JHEP {\bf 0909}, 038 (2009)
  [arXiv:0906.1461 [hep-ph]].
  
  \bibitem{He:2009ua} 
  X.~G.~He, S.~Oh, J.~Tandean and C.~C.~Wen,
  ``Large Mixing of Light and Heavy Neutrinos in Seesaw Models and the LHC,''
  Phys.\ Rev.\ D {\bf 80}, 073012 (2009)
  [arXiv:0907.1607 [hep-ph]].
  
  \bibitem{Adhikari:2010yt} 
  R.~Adhikari and A.~Raychaudhuri,
  ``Light neutrinos from massless texture and below TeV seesaw scale,''
  Phys.\ Rev.\ D {\bf 84}, 033002 (2011)
  [arXiv:1004.5111 [hep-ph]].
  
  \bibitem{Deppisch:2010fr} 
  F.~F.~Deppisch and A.~Pilaftsis,
  ``Lepton Flavour Violation and theta(13) in Minimal Resonant Leptogenesis,''
  Phys.\ Rev.\ D {\bf 83}, 076007 (2011)
  [arXiv:1012.1834 [hep-ph]].
  
  \bibitem{Mitra:2011qr} 
  M.~Mitra, G.~Senjanovic and F.~Vissani,
  ``Neutrinoless Double Beta Decay and Heavy Sterile Neutrinos,''
  Nucl.\ Phys.\ B {\bf 856}, 26 (2012)
  [arXiv:1108.0004 [hep-ph]].
  
  \bibitem{Dev:2013oxa} 
  C.~H.~Lee, P.~S.~B.~Dev and R.~N.~Mohapatra,
  ``Natural TeV-scale left-right seesaw mechanism for neutrinos and experimental tests,''
  Phys.\ Rev.\ D {\bf 88}, no. 9, 093010 (2013)
  [arXiv:1309.0774 [hep-ph]].
  
  \bibitem{Chattopadhyay:2017zvs} 
  P.~Chattopadhyay and K.~M.~Patel,
  ``Discrete symmetries for electroweak natural type-I seesaw mechanism,''
  Nucl.\ Phys.\ B {\bf 921}, 487 (2017)
  [arXiv:1703.09541 [hep-ph]].


 \bibitem{Dev:2012zg} 
  P.~S.~B.~Dev, R.~Franceschini and R.~N.~Mohapatra,
  ``Bounds on TeV Seesaw Models from LHC Higgs Data,''
  Phys.\ Rev.\ D {\bf 86}, 093010 (2012)
  [arXiv:1207.2756 [hep-ph]].
  
  \bibitem{Cely:2012bz} 
  C.~G.~Cely, A.~Ibarra, E.~Molinaro and S.~T.~Petcov,
  ``Higgs Decays in the Low Scale Type I See-Saw Model,''
  Phys.\ Lett.\ B {\bf 718}, 957 (2013)
  [arXiv:1208.3654 [hep-ph]].
  
  \bibitem{Hessler:2014ssa} 
  A.~G.~Hessler, A.~Ibarra, E.~Molinaro and S.~Vogl,
  ``Impact of the Higgs boson on the production of exotic particles at the LHC,''
  Phys.\ Rev.\ D {\bf 91}, no. 11, 115004 (2015)
  [arXiv:1408.0983 [hep-ph]].
  
  \bibitem{Das:2017zjc} 
  A.~Das, P.~S.~B.~Dev and C.~S.~Kim,
  ``Constraining Sterile Neutrinos from Precision Higgs Data,''
  Phys.\ Rev.\ D {\bf 95}, no. 11, 115013 (2017)
  [arXiv:1704.00880 [hep-ph]].
  
  \bibitem{Das:2017rsu} 
  A.~Das, Y.~Gao and T.~Kamon,
  ``Heavy neutrino search via semileptonic Higgs decay at the LHC,''
  Eur.\ Phys.\ J.\ C {\bf 79}, no. 5, 424 (2019)
  [arXiv:1704.00881 [hep-ph]].
  
  \bibitem{CMS:2018wxx} 
  CMS Collaboration,
  ``Searches for dijet resonances in pp collisions at $\sqrt{s}=13~\mathrm{TeV}$ using the 2016 and 2017 datasets,''
  CMS-PAS-EXO-17-026.


\bibitem{ATLAS:2019bov} 
  ATLAS Collaboration,
  ``Search for New Phenomena in Dijet Events using 139 fb$^{-1}$ of $pp$ collisions at $\sqrt{s}$ = 13TeV collected with the ATLAS Detector,''
  ATLAS-CONF-2019-007.


\bibitem{Okada:2016tci} 
  N.~Okada and S.~Okada,
  ``$Z^\prime$-portal right-handed neutrino dark matter in the minimal U(1)$_X$ extended Standard Model,''
  Phys.\ Rev.\ D {\bf 95}, no. 3, 035025 (2017)
  [arXiv:1611.02672 [hep-ph]].


 
\bibitem{Okada:2018ktp} 
  S.~Okada,
  ``$Z'$ Portal Dark Matter in the Minimal $B-L$ Model,''
  Adv.\ High Energy Phys.\  {\bf 2018}, 5340935 (2018)
  [arXiv:1803.06793 [hep-ph]].    



\bibitem{Okada:2017dqs} 
  N.~Okada, S.~Okada and D.~Raut,
  ``SU(5)$\times$U(1)$_X$ grand unification with minimal seesaw and $Z^\prime$-portal dark matter,''
  Phys.\ Lett.\ B {\bf 780}, 422 (2018)
  [arXiv:1712.05290 [hep-ph]].
  





\bibitem{Pumplin:2002vw} 
  J.~Pumplin, D.~R.~Stump, J.~Huston, H.~L.~Lai, P.~M.~Nadolsky and W.~K.~Tung,
  ``New generation of parton distributions with uncertainties from global QCD analysis,''
  JHEP {\bf 0207}, 012 (2002)
  [hep-ph/0201195].

 
 
\bibitem{Okada:2016gsh} 
  N.~Okada and S.~Okada,
  ``$Z^\prime_{BL}$ portal dark matter and LHC Run-2 results,''
  Phys.\ Rev.\ D {\bf 93}, no. 7, 075003 (2016)
  [arXiv:1601.07526 [hep-ph]].

 

\bibitem{Adam}
 A.~M.~Baldini {\it et al.} [MEG Collaboration],
  ``Search for the lepton flavour violating decay $\mu ^+ \rightarrow \mathrm {e}^+ \gamma $ with the full dataset of the MEG experiment,''
  Eur.\ Phys.\ J.\ C {\bf 76}, no. 8, 434 (2016)
  [arXiv:1605.05081 [hep-ex]].

\bibitem{Aubert}
B.~Aubert {\it et al.} [BaBar Collaboration],
  ``Searches for Lepton Flavor Violation in the Decays $tau^\pm \to e\pm \gamma$ and $tau^\pm \to  mu^\pm \gamma$,''
  Phys.\ Rev.\ Lett.\  {\bf 104}, 021802 (2010)
  [arXiv:0908.2381 [hep-ex]].
  
  
\bibitem{OLeary}
B.~O'Leary {\it et al.} [SuperB Collaboration],
  ``SuperB Progress Reports -- Physics,''
  arXiv:1008.1541 [hep-ex].
  
    
 \bibitem{deBlas:2013gla} 
  J.~de Blas,
  ``Electroweak limits on physics beyond the Standard Model,''
  EPJ Web Conf.\  {\bf 60}, 19008 (2013)
  [arXiv:1307.6173 [hep-ph]]. 
  
 \bibitem{delAguila:2008pw} 
  F.~del Aguila, J.~de Blas and M.~Perez-Victoria,
  ``Effects of new leptons in Electroweak Precision Data,''
  Phys.\ Rev.\ D {\bf 78}, 013010 (2008)
  [arXiv:0803.4008 [hep-ph]].  

 \bibitem{Akhmedov:2013hec} 
  E.~Akhmedov, A.~Kartavtsev, M.~Lindner, L.~Michaels and J.~Smirnov,
  ``Improving Electro-Weak Fits with TeV-scale Sterile Neutrinos,''
  JHEP {\bf 1305}, 081 (2013)
  [arXiv:1302.1872 [hep-ph]].    
  
  \bibitem{Esteban:2018azc} 
  I.~Esteban, M.~C.~Gonzalez-Garcia, A.~Hernandez-Cabezudo, M.~Maltoni and T.~Schwetz,
  ``Global analysis of three-flavour neutrino oscillations: synergies and tensions in the determination of $\theta_23, \delta_CP$, and the mass ordering,''
  JHEP {\bf 1901}, 106 (2019)
  [arXiv:1811.05487 [hep-ph]].

 \bibitem{Adamson:2016tbq} 
  P.~Adamson {\it et al.} [NOvA Collaboration],
  ``First measurement of electron neutrino appearance in NOvA,''
  Phys.\ Rev.\ Lett.\  {\bf 116}, no. 15, 151806 (2016)
  [arXiv:1601.05022 [hep-ex]].

 
 
 \bibitem{Abe:2017uxa} 
  K.~Abe {\it et al.} [T2K Collaboration],
  ``Combined Analysis of Neutrino and Antineutrino Oscillations at T2K,''
  Phys.\ Rev.\ Lett.\  {\bf 118}, no. 15, 151801 (2017)
  [arXiv:1701.00432 [hep-ex]].


  
   \bibitem{Aghanim:2018eyx} 
  N.~Aghanim {\it et al.} [Planck Collaboration],
  ``Planck 2018 results. VI. Cosmological parameters,''
  arXiv:1807.06209 [astro-ph.CO].  

 
 
\bibitem{Casas:2001sr} 
  J.~A.~Casas and A.~Ibarra,
  ``Oscillating neutrinos and muon$ \to e, \gamma$''
  Nucl.\ Phys.\ B {\bf 618}, 171 (2001)
  [hep-ph/0103065]. 



\bibitem{Lopez-Pavon:2015cga} 
  J.~Lopez-Pavon, E.~Molinaro and S.~T.~Petcov,
  ``Radiative Corrections to Light Neutrino Masses in Low Scale Type I Seesaw Scenarios and Neutrinoless Double Beta Decay,''
  JHEP {\bf 1511}, 030 (2015)
  [arXiv:1506.05296 [hep-ph]].
  
  \bibitem{Agostini:2018tnm} 
  M.~Agostini {\it et al.} [GERDA Collaboration],
  ``Improved Limit on Neutrinoless Double-$\beta$ Decay of $^{76}$Ge from GERDA Phase II,''
  Phys.\ Rev.\ Lett.\  {\bf 120}, no. 13, 132503 (2018)
  [arXiv:1803.11100 [nucl-ex]].

\bibitem{Dev:2013vxa} 
  P.~S.~B.~Dev, S.~Goswami, M.~Mitra and W.~Rodejohann,
  ``Constraining Neutrino Mass from Neutrinoless Double Beta Decay,''
  Phys.\ Rev.\ D {\bf 88}, 091301 (2013)
  [arXiv:1305.0056 [hep-ph]].

 
 \bibitem{DellOro:2016tmg} 
  S.~Dell'Oro, S.~Marcocci, M.~Viel and F.~Vissani,
  ``Neutrinoless double beta decay: 2015 review,''
  Adv.\ High Energy Phys.\  {\bf 2016}, 2162659 (2016)
  [arXiv:1601.07512 [hep-ph]].


\bibitem{Angrik:2005ep} 
  J.~Angrik {\it et al.} [KATRIN Collaboration],
  ``KATRIN design report 2004,''
  FZKA-7090.

\bibitem{Abazajian:2016yjj} 
  K.~N.~Abazajian {\it et al.} [CMB-S4 Collaboration],
  ``CMB-S4 Science Book, First Edition,''
  arXiv:1610.02743 [astro-ph.CO].



  
\end{thebibliography}
\end{document}